\documentclass[12pt,preprint]{aastex}
\shorttitle{O-type Stars in the MCs}
\shortauthors{Massey et al.}
\slugcomment{Submitted ApJ Jan 19, 2004; accepted Feb 25, 2004}

\begin{document}

\title{The Physical Properties and Effective Temperature Scale 
of O-type Stars as a Function of Metallicity. I. A Sample of 20
Stars in the Magellanic Clouds\altaffilmark{1}}

\author{Philip Massey\altaffilmark{2}}

\affil{Lowell Observatory, 1400 W. Mars Hill Road, Flagstaff, AZ 86001; Phil.Massey@lowell.edu}

\author{Fabio Bresolin and 
Rolf P. Kudritzki}
\affil{Institute for Astronomy, University of Hawaii, 2680 Woodlawn Drive, Honolulu, HI 96822; bresolin@ifa.hawaii.edu,kud@ifa.hawaii.edu}

\author{Joachim Puls and
A. W. A. Pauldrach}
\affil{Universit\"{a}ts-Sternwarte M\"{u}nchen, 
Scheinerstrasse 1, 81679, Munich, Germany; uh101aw@usm.uni-muenchen.de,uh10107@usm.uni-muenchen.de}

\altaffiltext{1}{Based on observations made with the NASA/ESA Hubble
Space Telescope, obtained at the Space Telescope Science Institute (STScI),
which is operated by the Association of Universities for Research
in Astronomy, Inc., under NASA contract NAS 5-26555.  These observations are
associated with programs 6417, 7739, 8633, and 9412. This paper also draws
heavily from data obtained from the data archive at STScI.}

\altaffiltext{2}{
Visiting astronomer, Cerro Tololo Inter-American Observatory (CTIO),
a division of the National Optical Astronomy Observatory, which is
operated by the Association of Universities for Research in Astronomy,
Inc., under cooperative agreement with the National Science Foundation.}

\begin{abstract}

We have obtained {\it HST} and ground-based observations of a sample of
20 O-type stars in the LMC and SMC, including six of the hottest
massive stars known (subtypes O2-3) in the R136 cluster.  In general,
these data include (a) the {\it HST} UV spectra in order to measure the
terminal velocities of the stellar winds, (b) high signal-to-noise,
blue-optical data where the primary temperature- and gravity-sensitive
photospheric lines are found, and (c) nebular-free H$\alpha$ profiles,
which provide the mass-loss rates.  We find that the older (FOS) {\it
HST} data of the R136 stars (which were obtained without the benefits
of sky measurements) suffered from significant nebular emission, which
would increase the derived mass-loss rates by factors of $\sim 3$, all
other factors being equal.  We also find several stars in the SMC
for which the N~III
$\lambda\lambda 4634,42$ and He~II$\lambda 4686$ emission ``f"
characteristics do not appear to follow the same pattern as in Galactic
stars. Since He~II emission is due to the stellar wind (which will be
weaker in SMC for stars of the same luminosity), while N~III emission
is a complex NLTE effect affected mostly by temperature, it would not
be surprising to find that these features do not correlate with each
other or with luminosity in SMC stars in the same was as they do 
in Galactic stars,
but theory does not provide a clean answer, and analysis of more stars
(both SMC and Galactic) are needed to resolve this issue.  The
line-blanketed non-LTE atmosphere code FASTWIND was then used to
determine the physical parameters of this sample of stars. We find good
agreement between the synthetic line profiles for the hydrogen, He~I,
and He~II lines in the majority of the stars we analyzed; the three
exceptions show evidence of being incipiently resolved spectroscopic
binaries or otherwise spectral composites.  One such system is
apparently an O3~V+O3~V eclipsing binary, and a follow-up radial
velocity study is planned to obtain Keplerian masses.  Although we did
not use them to constrain the fits, good agreement is also found for
the He~I $\lambda 3187$ and He~II $\lambda 3203$ lines in the near-UV,
which we plan to exploit in future studies.  Our effective temperatures
are compared to those recently obtained by Repolust, Puls \& Herrero
for a sample of Galactic stars using the same techniques.  We find that
the Magellanic Cloud sample is 3,000-4,000$^\circ$K hotter than their
Galactic counterparts for the early through mid-O's.  These higher
temperatures are the consequence of a decreased importance of wind
emission, wind blanketing, and metal-line blanketing at lower
metallicities.

\end{abstract}

\keywords{stars: early-type, stars: atmospheres, stars: fundamental parameters, stars: mass loss}

\section{Introduction}
\label{Sec-hist}

Translating the observed characteristics
of O and early B stars into physical properties has historically proven
to be be very challenging.   Because of their high effective temperatures,
most of their flux is in the far UV, where even spacecraft cannot
observe (due to interstellar extinction), 
as the peak of the flux distribution $F_\lambda$ will be below
the Lyman limit for stars hotter than about 32,000$^\circ$K.  This means
that the bolometric corrections are quite significant ($-1$ to $-4$~mags),
with a steep dependence on the effective temperature.  Thus getting
the effective temperature right is the key to determining the other
physical properties of a hot, massive star.

Since we are always observing the fluxes of these stars far on the tail of
the Rayleigh-Jeans distribution, the observed
colors of these stars show little sensitivity to effective temperature.
For instance, the Kurucz (1992) model atmospheres predict that a 
dwarf ($\log g=4.0$) with $T_{\rm eff}$=50,000$^\circ$K will have $(U-B)_o=-1.151$ and a $(B-V)_o=-0.326$, while a dwarf with 
$T_{\rm eff}$=40,000$^\circ$K will have a $(U-B)_o=-1.124$ and 
$(B-V)_o=-0.311$ (Massey 1998a). The difference in these colors is well 
within the error of most photometry, and yet the bolometric correction (BC)
is $-4.5$~mag for the former, and $-3.8$~mag for the latter, using the conversion of Vacca, Garmany, \& Shull (1996). This
uncertainty in the BC would result in an uncertainty
of 0.15~dex in the log of the mass, using the mass-luminosity relation
implied by the Schaller et al.\ (1992) $Z=0.02$ evolutionary tracks 
($L\sim M^2$; see discussion in Massey 1998a); 
i.e., 50$\cal M_\odot$ vs.~70$\cal M_\odot$.  
Reddening complicates the interpretation of the photometry further,
of course, and although reddening-free indices can be
constructed assuming an average reddening law, even an ``optimal"
reddening-free index using space-craft accessible UV photometry
(such as {\it F170W}) remains too degenerate to be useful (Massey 1998a).

In principle, spectroscopy should allow us to resolve this degeneracy
nicely, as the relative strengths of He~I and He~II (O stars) and Si~IV
and Si~III (early B supergiants) are quite sensitive to the effective
temperature (and only secondarily to the $\log g$), as confirmed by non-LTE
calculations (Auer \& Mihalas 1972; 
Kudritzki, Simon, \& Hamann 1983; 
Lennon et al.\ 1991).  Indeed, a difference of one spectral
subtype (O5~V to O6~V, say) roughly corresponds to a difference in
effective temperature of 0.02~dex in $\log T_{\rm eff}$, and a
corresponding difference of only 0.13~mag in the bolometric correction,
and hence 0.025~dex in the mass of the star ($50\cal M_\odot$
vs.~53$\cal M_\odot$) one would deduce from the mass-luminosity
relationship.

However, the absolute calibration of the relationship between
effective temperature and line strengths (or, equivalently, spectral
subtypes) requires reliable stellar atmosphere models.
Unfortunately, the physics of these stellar atmospheres
is quite complicated.
The strong lines are all formed under non-LTE conditions, as first
shown by Auer \& Mihalas (1972), and in addition stellar winds
provide a significant source of heating for the photosphere through
the backscattering of radiation
(Hummer 1982, Abbott \& Hummer 1985).
Each decade has seen an improvement in our understanding
of the physics of these stellar atmospheres, along with the development
the numerical techniques to include these effects in a model atmosphere.
We identify four stages in the evolution of these models and hence
in the effective temperature scale of O-type stars:

\begin{enumerate}
\item {\it The introduction of non-LTE.}  
The non-LTE models of Auer \& Mihalas
(1972) and Kudritzki (1975, 1976) were relatively simple by today's standards, 
in being plane-parallel and including no metals, but in fact required the
development of a number of innovative techniques in order to include the
effects of non-LTE in line formation. (See the review by Kudritzki \& Hummer
1990).  The Auer \& Mihalas (1972) models were the first to 
correctly reproduce the observed
line strengths of the helium and hydrogen spectral lines in O-type stars,
as was shown by Conti \& Alschuler (1971).  This led to the first modern
effective temperature scale for O stars, that of Conti (1973).  
This work was extended to the newly defined O3 spectral class
(Walborn 1971b) by Kudritzki
(1980) and Simon et al.\ (1983).

\item {\it The introduction of mass-loss.} 
Abbott \& Hummer (1985) showed that
the presence of stellar winds had a significant effect on the He~I/He~II line
ratios and hence on the effective temperatures deduced for O-stars.  Although
the photospheric lines  are formed in a nearly 
static part of the atmosphere, the 
scattering of radiation by the stellar winds back into the photosphere
results in substantial heating of the surface layers, an effect known as
``wind blanketing".   This effect is quite
significant, as a high (but realistic) mass-loss rate would result in
a 42,000$^\circ$K model matching an O3~V star, while the same model without
mass loss would match that of an O5.5~V star. In other words, the greater
the mass-loss rate, the cooler the effective temperature is for a given
spectral subtype.  Conti (1988) revised the effective temperature scale
to somewhat lower values, presumably to take this effect into account.

Of special note is the revised effective temperature scale of
Vacca et al.\ (1996), which has generally become the standard against which
other work is judged.  This scale was based on modeling drawn from
the contemporaneous literature, and as such represented the best that
the field had to offer at the time. Although it included {\it some} work that
was based on wind-blanketed models, the vast majority of the data
was not (e.g., Herrero et al.\ 1992), and so the derived temperature
scale was significantly higher than Conti (1988).

\item {\it The inclusion of hydrodynamics and metals.} 
The next decade saw substantial improvements in the modeling, 
with the introduction of 
spherical extension and a more sophisticated treatment of
the stellar wind as well as non-LTE treatment of the metal lines.
Rather than just using the mass-loss rates, the hydrodynamics of
the stellar winds in both the sub- and supersonic regions was
included. These improvements were pioneered by the stellar atmospheres
group in Munich (e.g., the {\it unified model atmospheres} concept; see
Gabler et al.\ 1989), which used them for the ``quantitative spectroscopy
of hot stars" (Kudritzki et al.\ 1989, Kudritzki \& Hummer 1990, 
Kudritzki 1991). (A similar but independent approach was taken by
Schaerer \& Schmutz 1994, who made the first attempt to include the
opacity of metal lines.)  Sellmaier et al.\ (1993) demonstrated that
in addition to the effect of wind-blanketing, the stellar winds produced
emission that partially filled in the He~I lines, which has a strong
effect on the effective temperatures as a function of the He~I/He~II line
ratios. 
Another seminal work from this period 
was Puls et al.\ (1996), who  
analyzed a large sample of Galactic and Magellanic Cloud O stars using
UV and optical data. The UV spectra were used
to determine the terminal velocities of the
stellar winds using the strong
resonance doublets N~V $\lambda\lambda 1239,43$, 
Si~IV $\lambda\lambda 1394,1403$, 
and C~IV $\lambda\lambda 1548,51$.
The mass-loss rates were then determined primarily by observations at
H$\alpha$, combined with an assumption (and occasional adjustment) of
$\beta$, a parameter that characterizes the steepness of the velocity
law in the stellar wind (Sec.~\ref{Sec-term}).  
The classical MK optical/blue region
(3800\AA\ to 4900\AA) still provides the primary diagnostics of
the surface gravity (from the hydrogen Balmer-line profiles) and
effective temperatures (from the He~I and He~II lines).  

\item {\it The full inclusion of line blanketing.}  A significant improvement
in hot star model atmospheres has been the inclusion of full line 
blanketing.  The CMFGEN code, described by Hillier \& Miller (1998)
and Hillier et al.\ (2003) is one such example.  Originally developed
and used for fitting emission-line features in the expanding atmospheres
of Wolf-Rayet stars, CMFGEN has been only recently used for the analysis
of absorption lines in O-type stars (i.e., Martins, Schaerer \& Hillier
2002, Crowther et al.\ 2002, Hillier et al.\ 2003, Bouret et al.\ 2003).
Although only a few stars have been fit, these studies suggest that the
effective temperature scale of Vacca et al.\ (1996) is too high\footnote{This would be expected in any event given the lack of wind-blanketed model fits
in the literature at the time of the Vacca et al.\ (1996) study.}.  (See
also discussion in Martins et al.\ 2002). Similarly,
WM-basic, a code developed by Pauldrach, Hoffmann, \& Lennon
(2001), includes full
line-blanketing, and its use by Bianchi \& Garcia (2002) also suggests
that a lowering of the effective temperature scale is in order.
(WM-basic lacks the Stark broadening and the co-moving frame treatment
needed to compute useful synthetic
spectra of the quasi-photospheric hydrogen and helium lines, 
but is very useful to
fitting the metal lines found in the UV spectrum.)  
FASTWIND (``Fast Analysis of STellar atmospheres with WINDS"), first
described by
Santolaya-Rey, Puls, \& Herrero (1997), has now been modified 
to include an approximate---but highly realistic---treatment of line-blocking,
with a similar effect on the effective 
temperatures 
(Herrero, Puls, \& Najarro
2002, 
Repolust, Puls, \& Herrero 2004).\footnote{The use of 
an {\it approximate} treatment of line-blanketing and
blocking is necessitated by the need for reasonable computational times
for a model.  We note for comparison that a single run of CMFGEN, with
its more rigorous treatment, requires 9 hours on a 1.3-GHz Pentium IV
processor, according to Smith, Norris, \& Crowther (2002).   By contrast,
a single run of FASTWIND requires less than 7 minutes on the slower
750-Mhz SparcIII machine we used in this study.
In general, something like 10 to 30
models are needed to fit a star. Comparisons of the flux distribution
of FASTWIND with those of WM-basic 
(Repolust et al.\ 2004) and CMFGEN (Herrero et al.\ 2002) have
so far shown
very good agreement.}
 
\end{enumerate}

In this series of papers, we will use new observational data with the latest
generation of model atmosphere code to derive a new effective temperature
scale for O and early B-type stars, exploring for the first time
 the effect that metallicity plays on the effective temperature
scale (and other derived physical parameters) of these stars.  This
is crucial for many astrophysical applications, such as deriving
the initial mass function from H-R diagrams (Massey 2003), or in
modeling expanding shells, super-bubbles, and normal H~II regions,
where having an accurate census of the ionizing flux and amount of
mechanical energy being supplied to the region by stars is crucial
(Oey \& Kennicutt 1997).   The series will eventually encompass hot stars
in the SMC, LMC, Milky Way, and the Andromeda Galaxy, which span a factor
of  6.8 in metallicities, at least as measured by the oxygen abundances
(see Massey 2003 and references therein).  Here we begin by studying a
sample of hot stars in the Magellanic Clouds.  Analysis of  a second
sample of Magellanic Cloud stars is currently underway, and will be
published shortly.

\section{Observations and Reductions}

We list in Table~\ref{tab:stars} the identifications and spectral types of
the sample of stars we consider here.  As previously noted, our modeling
requires observations taken in three spectral regions: (1) the 1200-1900\AA\
region where the strong stellar wind resonance lines are found, in order
to determine the terminal velocity of the wind; 
(2) the blue-optical region of the spectrum, where the strong Balmer hydrogen
lines determine the effective gravity of the star, and the He~I and He~II
line strengths determine the effective temperature; and 
(3) the H$\alpha$ line
profile, used to determine the mass-loss rate and the value for $\beta$,
a measure of the steepness of the stellar wind velocity law
(e.g., Puls et al.\ 1996). 
In principle, values
for all three of these are interdependent, but in practice, a reliable
value for the terminal velocity can be determined by simple fitting with
very approximate values for the other stellar parameters.  Determining of
the effective temperature, surface gravity, and mass-loss rate does, however,
require simultaneous treatment.

\subsection{Photometry}

One of the necessary input parameters for the modeling is an accurate knowledge
of the star's absolute visual magnitude $M_V$ in order to constrain the
stellar radius.  Fortunately, with stars in the
Magellanic Clouds there is no uncertainty about the relative distances,
and even the absolute distances are now known to reasonable accuracy  
(van den Bergh 2000).  However, our experience is that photoelectric
photometry of stars in the Magellanic Clouds is simply not as good as
CCD photometry, as the large apertures used in the former often allowed
contamination by nearby stars and/or nebular emission.  Photometry with
CCDs allows local sky subtraction, and the issues of crowding can be
dealt with either by the use of small digital apertures or, in extreme
cases, point spread function fitting techniques.  Accordingly, Massey (2002)
obtained {\it UBV} photometry of nearly all of the stars in our target list
(outside of the R136 cluster) and, coincidentally, of an additional 264,600 
stars.  We list the catalog number in Table~\ref{tab:stars}; not all of
the stars have cross-references in Massey (2002) Tables 4 and 6.  The
color excesses at $U-B$ and $B-V$ were determined using the intrinsic
colors expected for each spectral type (Massey 1998b), and the results
averaged, with the assumption that $E(U-B)=0.72 \times E(B-V)$.  For the
R136 cluster, the WFPC2 photometry of Hunter et al.\ (1997) was used
by Massey \& Hunter (1998) to derive $M_V$, and we adopt these values
here.

\subsection{Spectroscopy}
\label{Sec-data}

In Table~\ref{tab:journal} we list all of the data directly used in our
study.  These were obtained from both {\it HST} and the 
CTIO 4-m telescope, and cover the UV, H$\alpha$, and optical/blue regions.

\subsubsection{Ultraviolet}

For all of our program stars, we used  
{\it HST} to obtain spectra in the UV region, where the major
stellar wind resonance lines  N~V$\lambda\lambda 1239,43$, 
Si~IV $\lambda\lambda 1394, 1403$, and 
C~IV $\lambda\lambda 1548,51$ are located.
Our data came both from archival programs and our own program.

For the ultraviolet observations of the
R136 stars we used the archival {\it HST} observations made
with the Goddard High Resolution Spectrograph (GHRS) 
by S. Heap 
(PI) under programs GO-5297 (R136-20=R136a5) and GO-6018 (R136-24=R136a7,
R136-36=R136a-608, R136-40=R136a-535, R136-47=R136a-602, and
R136-55=R136a-551.  The data were obtained in 1994 and 1996, as
shown in Table~\ref{tab:journal}, and hence were post-COSTAR.
These UV data were the ones used by
de Koter, Heap, \& Hubeny (1997, 1998), and were
obtained with the G140L grating centered at two wavelength
settings, 1300\AA\ and 1610\AA, and covered the spectral regions
1160-1450\AA\ and 1460-1750\AA, respectively.  The resolution
of the GHRS data is 0.6\AA.  The observations were made through the
`small science aperture", which was 0.22 by 0.22 arcsec in size.

For the ultraviolet observations of
most of the SMC stars, we used the FOS observations made by C. Robert
(PI) under program GO-5444.  These data were obtained in 1994 and 1995, and
were mentioned by Robert (1999) and included in Leitherer et al.\ (2001). 
The data were obtained with the G130H grating covering 1140-1606\AA,
with a resolution of 1.0\AA.
The data were taken with a 3.7 by 3.7 arcsec aperture.  

For AV~296 in the SMC, and the three LMC stars outside of the R136
cluster (LH64-64, LH81:W28-5, 
and LH101:W3-24), we obtained our own 
UV observations using STIS/FUV
under program GO-8633.(Massey, PI).  The 
G140L grating was centered at 1425\AA\ for a wavelength coverage of
1150-1736\AA, with a resolution of 0.9\AA.  
The objects were observed through a 0.2 by 0.2 arcsec aperture.
In addition, we obtained an observation with the same setup
for the SMC star AV~26 (previously
observed by Robert 1999 with the FOS) with STIS/FUV under program
GO-9412 (Massey, PI). This was intended to serve as a self-consistency
check for the terminal velocities obtained with the FOS and STIS.

\subsubsection{Optical}
\label{Sec-optical}

For all of the R136 stars we obtained optical data 
(both blue and H$\alpha$) using {\it HST}.  For
the rest of the stars, we obtained optical data primarily with the CTIO 4-m
telescope.  However, in a few cases, where we feared nebular 
contamination of the H$\alpha$ profile, or in cases
where the stars were very crowded, we supplemented the ground-based
data with {\it HST}.

For the optical (blue and H$\alpha$)
observations of the R136 stars, we obtained spectra with high
signal-to-noise ratios (SNRs) using STIS/CCD.
These data were obtained under GO-7739 during 1998 Feb 3-5.  These
stars had all had previous optical FOS
observations by de Koter et al.\ (1997, 1998) and/or 
Massey \& Hunter (1998), 
but we knew from our
own spectroscopy of R136 stars with the FOS (Massey \& Hunter 1998) that
the maximum SNR achievable with the FOS
was quite limited 
(typically 50 per quarter-diode).
In addition, we expected that nebular
contamination of the H$\alpha$ profile was likely given the lack of
sky subtraction.  This indeed proved to be the case, as we show in
Sec.~\ref{Sec-compare}. 
STIS offered the advantage of a higher SNR plus
the ability to subtract nebular emission thanks to the long slit and
two-dimensional format of the detector. 
Nevertheless, the FOS observations
are still useful in that they contain spectral lines that are not covered
in our STIS observations, particularly He~II $\lambda 4686$, and we
make use of the Massey \& Hunter (1998) observations for this line. 
Our blue R136 STIS observations were made with the G430M grating 
centered at 4451\AA, and covering the wavelength range
4310\AA\ to 4590\AA, chosen to include H$\gamma$, He~I$\lambda 4387$,
He~I $\lambda 4471$, and He~II $\lambda 4542$.
The spectral resolution was 0.4\AA.  We chose our exposure times to
achieve a SNR of 100 per spectral resolution element.
The H$\alpha$ observations were obtained
with the G750M grating, centered at 6581\AA, and covering the wavelength
range 6300\AA\ to 6850, with a spectral resolution of 0.84\AA.  Our exposure
times were chosen to achieve a SNR of 50
per spectral resolution element.  The observations were made with the
0.2$\times$52 arcsec slit.  The spatial sampling was 0.05 arcsec~pixel$^{-1}$
along the slit.  The FOS observations of Massey \& Hunter (1998) were made
with a 0.26 arcsec diameter aperture, and covered the 3250\AA\ to 4820\AA\
region with 3\AA\ resolution.  Those data were obtained under GO-6417 (Massey, PI).

The optical observations for the other stars were 
mostly
obtained during a 5 night run on the CTIO 4-m Blanco
telescope with the RC Spectrograph during 3-7 Jan 1999.  Grating KPGL-D
was used in 2nd order with a CuSO$_4$ blocking filter for observations
in the blue (3750-4900\AA).  The spectral resolution with the 200$\mu$m
(1.25 arcsec) slit was 1.4\AA\ (3.8 pixels) on the Loral 3Kx1K 15$\mu$m CCD
behind the blue air Schmidt camera.  The slit length was approximately
5 arcminutes, with a spatial sampling of 0.5 arcsec pixel$^{-1}$ (i.e.,
ten times coarser than with STIS/CCD).
For observations at H$\alpha$ we switched to 1st order and observed from
5400 to 7800\AA\ with a GG420 filter to block 2nd order blue.  The spectral
resolution for the H$\alpha$ observations is 2.8\AA.
In general, conditions were excellent during the run.  The stars were 
well exposed, and by considerable efforts at flat-fielding, we were able
to preserve the high SNR of the data, typically 400 to 500 per
spectral resolution element.   This was a particular challenge in the
blue, where exposures of many hours of the dome flat barely achieved
a SNR of 100 at 4000\AA.  Instead, we combined 27 
30-sec projector flats to achieve a very high SNR flat suitable
for removing the pixel-to-pixel variations, but whose 
overall illumination function did
not match the sky.  We corrected the projector flats by using the
average of 9 1200-second dome-flat (``Punto Blanco") exposures.  Each set
of flats was combined with deviant pixel rejection.   At H$\alpha$ it was
practical to simply use a series of dome flats to generate an adequate 
flat-field exposure.

We also have H$\alpha$ {\it HST} STIS/CCD observations for two of the
non-R136 stars.
LH101:W3-24 is located in a
region of strong nebulosity, and we were unable to obtain a successful
observation from the ground at H$\alpha$, and so we observed it with a
narrow (0.2 arcsec) slit with {\it HST} using the G750M with the same
setup as used for the R136 stars.  For AV~26, the same star for which
we obtained both STIS and FOS UV observations as a consistency check
for the terminal velocities, we also used an observation at H$\alpha$ as
an additional check against our ground-based observation in a
``typical" case.  Both were observed as part of program GO-9412
(Massey, PI).  In order to fill up the remainder of the orbit for each
of these two stars, we also observed them in the near-UV region of the
spectrum, where the He~I $\lambda 3187$ and He~II $\lambda 3203$ lines
are located (see Morrison 1975).  These data were obtained
with the G430M grating centered at 3165\AA, covering 3020-3300\AA, with
a resolution of 0.4\AA.  All of the STIS/CCD observations were done in
at least two exposures (``cr-split"); in addition, all but the near-UV
observation of LH101:W3-24 were dithered along the slit at three
positions in order to increase the SNR.

Three of the program stars (AV~378, AV~396, and AV~451) had no observations
at H$\alpha$ due to on-going problems with the CTIO 4-m control system
during the run.  We have included these stars in the analysis despite this,
making assumptions outlined in Sec.~\ref{Sec-analysis}

\subsubsection{Reductions}

The data reduction proceeded as follows.  For the {\it HST} UV observations,
all of which were obtained through small apertures, we accepted the {\it HST}
CALSTIS pipeline versions of the reductions.  
For the {\it HST} optical (blue and H$\alpha$) 
long-slit STIS/CCD observations, we re-reduced the data 
{\it ab initio}, using the recommended flat, dark, and bias frames.  We have
found that by using the standard spectral reduction algorithms in 
IRAF\footnote{IRAF is distributed by the National Optical Astronomy Observatories,
    which are operated by the Association of Universities for Research
    in Astronomy, Inc., under cooperative agreement with the National
    Science Foundation.}
(which include optimal extraction and profile-based pixel rejection) we
can generally achieve a SNR that is significantly better than that 
produced by the pipeline.  As others may also benefit from our experience,
we show the difference in Fig.~\ref{fig:diff}.  The SNR of the standard
CALSTIS pipeline is worse than that of the re-reduced IRAF data for
several incremental reasons.  First, good data are lost in the CALSTIS
cosmic-ray rejection when the two halves of a ``cosmic-ray split"
image are combined. The same operation using IRAF can readily be made less aggressive. (Admittedly one could achieve the same result by
re-combining the images using the STSDAS implementation of CALSTIS
with similar adjustment of parameters.)
Second, some of the SNR is lost because CALSTIS fails to do the spectral
extraction (summing over the spatial profile) using
an optimal extraction routine.  
With optimal extraction (Horne 1988; Valdes 1992), 
each point in
the profile is summed using a weight that is inversely proportional to
the square of the sigma expected on the basis of the read-noise and
signal level. Third, IRAF uses the shape of the profile to reject
highly deviant pixels (Valdes 1992).  As shown in the figure, this is quite effective
at reducing the effects of cosmic rays and hot pixels.  For shorter exposures,
and higher SNR spectra, we found less of a difference, but invariably the
IRAF spectrum was to be preferred. The same IRAF extraction routines were used
for reducing our CTIO 4-m data.

\subsection{A Comparison of Our H$\alpha$ Spectra with Previous Studies}
\label{Sec-compare}

Earlier, we expressed our concern that observations at H$\alpha$ could give
erroneous results in the absence of sky (nebular) subtraction, particularly
in dense H~II regions such as the one in which the R136 cluster is situated.
Contamination by nebular H$\alpha$ would invariably lead to spuriously
large mass-loss rates.  The H$\alpha$ data used in this study have all come
from STIS long-slit observations which allow good sky/nebular subtraction.
Earlier studies by de Koter et al.\ (1997, 1998) of R136, 
though, were forced to rely upon observations made with the FOS and a 
single aperture.
Somewhat suggestively, these studies found that the mass-loss rates
of the R136 stars were considerably higher at a given luminosity than
that of other O stars that had been studied by Puls et al.\ (1996).  
This difference could be real, or it could be due to the different model
atmospheres used to derive the mass loss rates---or it could potentially 
be due to the observations
themselves.  How well do the older FOS data compare to ours?

The results are shown in Fig.~\ref{fig:heap}.  Our STIS (nebular-subtracted)
spectra are shown in green, with the FOS (no sky subtraction) spectra used
by de Koter et al.\ (1998) shown in red.  In {\it every} case there is
additional emission present in the latter.  The fainter stars (R136-040,
R136-047, and R136-055) show the larger effect, as would be expected in
the case of nebular contamination.  For these stars, the FOS spectra
fail to even detect the underlying absorption feature, nicely revealed
by our sky-subtracted STIS spectra.
The de Koter et al.\ (1998) code did prove its
flexibility in being able to match these (spurious) emission features
in their stellar modeling.  However, in these cases analysis of the FOS 
would of necessity lead to erroneously
high mass-loss rates. 

How much of an error would this extra emission introduce?  We can provide
an approximate answer by taking the parameters we derive for these stars
in Sec.~\ref{Sec-analysis}
and simply increasing the mass-loss rates until we have emission that
approximately mimics that of the FOS data.  We find that mass-loss rates
of 2-4 times what we derive from the STIS data would be needed to 
approximate the FOS data. (In point of fact we could not get our H$\alpha$
profiles to match those of the FOS data.)  The mass-loss rates we eventually
derive for the R136 stars are in fact not that different than those of
de Koter et al.\ (1998) (and in some cases are actually higher!) but this
is due to a combination of factors, primarily the much higher temperatures
we find from our greater SNR optical data.

Finally, in Fig.~\ref{fig:AV26comp} we compare our {\it HST} STIS H$\alpha$
spectrum of AV~26 to that obtained with the CTIO 4-m.  The agreement is
quite good, despite the 3.5 times worse resolution of the
ground-based data (see Table~2).

\section{Analysis}
\label{Sec-analysis}

\subsection{Terminal Velocities}
\label{Sec-term}

The first step in our analysis was fitting the UV lines to determine the 
stellar wind terminal velocities. 
Terminal velocities were measured from radiative transfer fits of the
P-Cygni profile of the C~IV $\lambda 1550$ 
doublet.  Of the other important
resonance lines present in our UV spectra, the shortwards
profile of N~V $\lambda 
1240$ is
often contaminated by strong interstellar Ly $\alpha$ absorption, making
measurements of the terminal velocities from this line very uncertain in
most cases. 
Si~IV $\lambda 1400$, on the other hand, was often weak, and thus would
not allow us precise constraints on $v_\infty$. We have followed the
fitting technique described by Haser (1995) (see also Haser et al.\ 1995),
based on the SEI method (cf.\ Lamers, Cerruti-Sola, \& Perinotto 1987). This method has been used
in more recent investigations of the UV spectra of Galactic and
extragalactic O and B stars, including the HST/STIS work by Herrero et al.\ 
(2001), Urbaneja et al.\ (2002), and Bresolin et al.\ (2002), to which we
refer the reader for details. We allow for a radially increasing 
turbulent velocity law
in the stellar winds, which is described by the usual $\beta$ parameterization
($\beta\simeq 0.8$ for O stars):
$$ \frac{v(x)}{v_\infty} = \left( 1 -
\frac{b}{x} \right)^\beta $$
where $x= r/R$ is the radial coordinate normalized to the stellar
radius $R$, 
and $b$ fixes the velocity of the inner boundary of the wind to
$V(R)$, a value that is of order of the sound speed (Kudritzki \& Puls
2000), i.e.,
$$b=1-\left(\frac{V(R)}{v_\infty} \right)^{1/\beta}.$$
The best fit to the shortward line profile, which is mostly sensitive
to the adopted terminal wind velocity, provided the results summarized in
Table~\ref{tab:termvels}.  
In most cases the uncertainties in $v_\infty$ are of the order
of 50 to 100 km s$^{-1}$. Larger uncertainties (up to $\sim 200$ km s$^{-1}$)
 are
estimated for those stars having weak C IV lines (noted by the colons
appended to their $v_\infty$ value in Table~\ref{tab:termvels}), 
and for which we have
also relied on the Si IV lines for estimating $v_\infty$.  Typically
the maximum turbulent velocity is 9\% of $v_\infty$, with values
ranging from 3\% to 14\%.

In Table~\ref{tab:termvels} we also compare our terminal velocities
to those of Prinja \& Crowther (1998) and de Koter et al.\ (1998) for the
R136 stars in common.  In considering this comparison, it is worth noting
that the terminal velocities were measured {\it  from the identical data}.
In other words, the differences between these measurements are purely
due to technique.  To the best of our knowledge, this is the first time
such a comparison has been carefully performed with the same data.
We see that the agreement is fairly good, a few hundred
km~s$^{-1}$, consistent with our estimate of our own uncertainty, but far
greater than 
might be naively inferred by the precision with which these measurements
are occasionally published.
Perhaps fortuitously, our second measurement of the terminal velocity of
AV~26 (which was observed both with the FOS and STIS/FUV) agreed to the
best of our measuring accuracy.  This gives us confidence that the different
instrumentation does not introduce much of a bias.  We show the CIV lines
from the two observations in Fig.~\ref{fig:UV}, along with our modeling
of the terminal velocity, as an example.

\subsection{Model Fits}
\label{Sec-fits}

The stellar atmosphere code generates synthetic line profiles given the
inputs of effective temperature $T_{\rm eff}$, surface gravity $g$,
the stellar radius $R$, the mass-loss rate $\dot{M}$,
the stellar wind terminal velocity $v_\infty$, the He/H number ratio, 
and the metallicity $Z/Z_\odot$.
For a given model, the parameters $q(\infty$), $q_0$, and $\gamma$ of the
non-LTE Hopf function (Santolaya-Rey et al.\ 1997, Mihalas 1978) 
must be adjusted until flux conservation ($<2$\%) 
is achieved; in practice, this
requires several runs.  Good starting points for the Hopf parameters were
found by interpolating of successful values from previous runs of similar
input values.

In fitting a star, we adopted the terminal velocities determined in the
previous section, and assumed a metallicity $Z/Z_\odot$ of 0.2 for the 
SMC stars and 0.5 for the LMC stars\footnote{The 
values for the metallicities are certainly
arguable: as Westerlund (1997) notes, the relative abundances of the  
interstellar medium in the SMC, LMC, {\it and} the nearby regions of
the Milky Way are non-solar.  Based primarily on the work of Russell \& Dopita
(1990), Westerlund (1997) argues that the ``average" metal abundance is
0.6~dex and 0.2~dex lower in the ISM of the SMC and LMC than in the solar
neighborhood; see also Garnett (1999). 
The gas in the solar neighborhood is perhaps 0.1~dex lower
than that of the Sun (Shaver et al.\ 1973, Cameron 1982; Table 8 of
Russell \& Dopita 1990), although recent revisions in the solar abundances
may suggest otherwise (Asplund 2003).}
A He/H number ratio of 0.10 was
adopted, and adjusted if needed. For several stars the helium lines
produced by the models
were too weak compared to the theoretical hydrogen lines, and we had to
increase the He/H ratio, as described below.
If the He/H ratio was increased, the relative mass fractions of the 
other elements
were retained. Although this is not quite right (as some elements, such
as  nitrogen,  would likely also increase in abundance, while
carbon and oxygen would decrease) 
it does preserve the overall fraction of metals,
particularly the unprocessed Fe group elements, which are most important
in the blanketing.  
An examination of the evolution of the metallicity in the cores of massive
stars suggests that this is a good approximation: although the relative
proportion of elements changes during core H-burning, the overall fraction
of the mass of the star that is in metals changes very little until He-burning
products are produced\footnote{Compute the value $Z=1-(Y_c+X_c)$ as a function
of time in the models of Schaller et al.\ (1993), for example.  One will
find for the $120\cal M_\odot$ (their Table~1)
that metallicity {\it in the core}
changes insignificantly
from 0.0082 to 0.0077 during core H-burning.  The same numbers
are found for a star of $20\cal M_\odot$, as found in their Table~6.}.
A starting value for the mass-loss rate was estimated
by adopting the bolometric correction
based on the spectral type using the Vacca et al.\ (1996) calibration
and adding this
to the absolute visual magnitude (Table~\ref{tab:stars}) in order to 
get a crude approximation of the bolometric luminosity L; a mass-loss
rate based upon Puls et al.\ (1996) and scaled appropriately by the
metallicity was then used for a 
first approximation. Following Repolust et al.\ (2004), we adopted
a micro-turbulence 
velocity of 10 km s$^{-1}$ for the models with 
effective temperatures of 36,000$^\circ$K and below, and 0 km s${-1}$ for
hotter stars.
A grid of 3-9 models using ``reasonable" values for
the effective temperature and surface gravity (based upon the spectral type
and the Vacca et al.\ 1996 scale) was then run.  The initial starting value
for the stellar radius was based upon the effective temperature and the
approximate bolometric luminosity (using the relationship between the
bolometric correction and $T_{\rm eff}$ of Vacca et al.\ 1996).  For each
grid point the true radius was then computed using the model flux and
effective temperature.  If the input and derived radius differed by more
than 1\%, the input value was adjusted and the grid point recomputed.

After the first series of models is run, a comparison is made by eye between
the synthetic spectra from the models and the observed spectrum. For this,
both a radial velocity and rotational speed $v \sin i$ must be adopted;
these were determined prior to the modeling by examining weak lines in the
optical spectrum.  In the modeling occasionally this initial measurement
had to be slightly refined (by 10\%) for a given star.  In most cases we
found $v \sin i \sim $110 or 120 km s$^{-1}$, comparable to the instrumental
resolution of the ground-based data, so these values should not be
over-interpreted.
In general, the wings of the Balmer hydrogen lines H$\gamma$
and H$\delta$ are the primary diagnostics of whether the surface gravity is
about right, while the relative strengths of the He~I and He~II lines provide
the greatest sensitivity to effective temperature.  The H$\alpha$ profile
provides the key diagnostic of the mass-loss rate.

For most of the stars, we achieved excellent agreement between the 
model synthetic spectra and the observed
spectra.  However, for three of the
stars, no adequate fits could be achieved.
We believe this is only to be expected: the frequency of close binary O-type
stars in the Milky Way is about 35\% (Garmany, Conti, \& Massey 1980); in
most of these systems, the mass ratio is near unity with the companion another
O-type star.  If the binary frequency is similar in the LMC and SMC (a subject
for which there is little data to argue one way or the other), then we
would naively expect about one-third of our sample to consist of stars with
composite spectra.  One or two additional stars are likely
unrecognized composites. Subsequent to our beginning this study,
one of the stars (R136-024) for which no good fit could be found
was shown to have light variations typical
of eclipses (Massey, Penny, \& Vukovich 2002), further supporting this interpretation.

Repolust et al.\ (2004) discuss a long-standing problem with stellar 
atmosphere
codes (see Voels, Bohannan, \& Abbott 1989) namely
that He~I $\lambda 4471$ (which is
one of the principal spectral classification lines) synthetic spectrum
is generally {\it weaker} than that observed in the cooler O-type giants
and supergiants among their Galactic sample.  This problem is not 
understood to date, although we are continuing to investigate various
causes.
Here we can say that we find the
same problem as reported by Repolust et al.\ (2004) for Galactic stars
in our lower-metallicity Magellanic Cloud sample. Since the problem seems
to only affect the lower-gravity stars cooler than O6, this is in
practice not an issue, as for these stars we can rely upon other He~I
lines which are apparently not affected by this problem, most notably
He~I $\lambda 4387$. 

Repolust et al.\ (2004) consider the 
possibility that the He~I $\lambda 4471$
problem might be resolved by including a more consistent calculation of
the temperature structure, including that of the outer part of the stellar
wind.
We originally used the identical version of FASTWIND as used by Repolust
et al.\ (2004) for making the model fits.  Near the end of our study,
work on an improved version was completed. This version now includes
a self-consistent temperature stratification, in which the equation of
thermal balance of electrons is used to derive the temperature structure,
except in the innermost region of the photosphere, where a flux correction
method  is used (see
Kubat, Puls, \&
Pauldrach 1999). In addition, the new version includes a more
extensive 
line list,  leading to better fluxes in the UV.
We re-ran this improved version on all of our models, and carefully
compared the results.  Only the fits for the stars with very high 
mass-loss rates (R136-020 and R136-036) were significantly 
affected, as we expected.
In those cases the new derived temperatures and mass-loss rates were
somewhat lower than those produced with the earlier version.  No
improvement was seen in the agreement for the He~I $\lambda 4471$ line
for the problem stars.  The numbers and fits given here are all from
this newer version; comparison with the Galactic sample analyzed
by Repolust et al.\ (2004) should be valid, as the mass-loss rates are
comfortably low, except for HD~93129A, which is a spectroscopic binary
and excluded from consideration below.

Given that there is a certain amount of subjectivity in determining the
``best fit", what are reasonable uncertainties on the fitted parameters?
In general, we found that we could determine effective temperatures with
an uncertainty of 1,000$^\circ$K, and $\log g$ to 0.1~dex.
In the cases where we had to increase the He/H ratio from 0.10 in order
to get a good fit, we generally were confident of our value to $\pm 0.05$,
except for LH64-16, for which we find a very high He/H ratio, with an
uncertainty that is about 0.1.
The 
uncertainty in $\dot{M}$ is generally less than 20\%, although we caution
that our models lack stellar wind clumping, and we expect that this will affect
the derived mass-loss rates for stars with H$\alpha$ in emission, but
with minimal impact on the majority of our stars, for which H$\alpha$ is
in absorption (see Repolust et al.\ 2004).
For stars with the lowest mass-loss rates 
($<3\times10^{-7} M_\odot$ yr$^{-1}$) we include both the value
used in the fit and the upper limit (under ``Comments") in Table~4.

\subsection{Spectral Classification Issue at Low Metallicity.}

In general, we followed the precepts of the Walborn \& Fitzpatrick (1990)
atlas in (re)classifying the stars. The basis
for the spectral subtype is primarily the ratio of the He~I to He~II ratios, 
particularly that of He~I $\lambda 4471$ and He~II $\lambda 4542$ (cf.
Conti \& Alschuler 1971).  For stars that would have been classically 
classified as ``O3", we used the criteria given by Walborn et al.\ (2002a)
to classify these stars as either O2 or O3 when we could.
In addition to assigning a spectral subtype
on the basis of the overall {\it appearance} of the spectrum (based upon
comparision
to the Walborn \& Fitzpatrick 1990 atlas), we also {\it measured} the
equivalent widths of the He~I $\lambda 4471$ and He~II $\lambda 4542$
lines, and compared the ratio to that used to define the spectral subtypes
by Conti \& Alschuler 1971).  Below we quote the quantity 
$\log W'=\log W(4471)-
\log W(4542)$, where $W$ is the equivalent width.  In practice, there were
no differences between the two methods.

Assigning the luminosity class to these stars is slightly trickier.
The amount of emission in N~III $\lambda\lambda 4634, 42$ and He~II
$\lambda 4686$ results in an ``f" designation, with the spectroscopic
description ``((f))" referring to weak N~III emission and strong He~II
$\lambda 4686$ absorption.  The description ``(f)" refers to N~III
emission with partially filled in He~II absorption, while ``f" refers
to both features being in emission.  In the Milky Way the ``f"s are
invariably supergiants (i.e., "If" stars), while the ``(f)" are giants
(i.e., ``III(f)"), and the ``((f))" are invariably dwarfs (i.e.,
``V((f))").  But at low metallicity, such as found in the SMC, this
shouldn't necessarily follow. 
The amount of emission in He~II $\lambda 4686$ 
is a function of what is  happening in the stellar wind 
(Klein \& Castor 1978, Gabler et al.\ 1989).  
The physics involved in the formation of this line is not as simple as
that which leads to, say, H$\alpha$. The formation of H$\alpha$ will be 
largely unaffected by blocking in the EUV, while HeII $\lambda 4686$
may be strongly affected, due to the importance of the HeII $\lambda
303$ resonance line in its formation. The difference in metallicity between
the Milky Way and the SMC will mean that two stars of the same spectal 
subtype will have different effective temperatures, different wind densities, 
and will experience different amounts of EUV blocking.  These competing
effects make it hard to predict from first principles whether there will
be {\it more} emission, or {\it less} emission for stars
of the same spectral subtype in the SMC and Milky Way.  Similarly, the
emission in N~III $\lambda\lambda 4634, 42$ is a complex NLTE effect,
and its size to large extent dependent upon the effective temperature
(Mihalas \& Hummer 1973; Taresch et al.\ 1997).  Without any stellar
wind, He~II $\lambda 4686$ would be in absorption in stars of all
luminosity classes, while N~III would still show emission.  Thus, it is
not unreasonable to expect that the ``f" properties of a star may not
follow our Galactic prejudices as we look at O stars at lower
metallicities.  We will therefore call attention to the $M_V$ when
discussing the stellar classifications, as abhorrent as this practice
may seem to classification purists.

Of course, one can always argue that any particular star with a discrepantly
high visual luminosity may be a spectroscopic binary.  At the distances
of the Magellanic Clouds such blends might not even be revealed by
radial velocity motion, given that 1" projects to 0.24 pc (LMC) or
0.29 pc (SMC).  However, below we find several examples 
(AV~14, AV~26, AV~75, and possibly AV~469) where the
``f" characteristics are not totally consistent with the star's $M_V$ in
the SMC, but none in the LMC.  To explain these away as binaries would
require our SMC sample to be biased towards unresolve multiple systems
in a way that the LMC sample is not.
While not impossible, the alternative explanation that we are
seeing a metallicity effect on the ``f" characteristics generally used to
define the luminosity class, would appear to us as more attractive.

\subsection{Comments on Individual Stars}

In this section we discuss the derivation of the spectral types, and
present the results of our model fits.  The physical parameters are
given in Table~4.  We include in that table the so-called
``spectroscopic mass", $M = g/g_\odot \times R^2$.  In the figures showing
the spectra, occasional bad columns have been removed by linear interpolation.

\subsubsection{SMC}

{\it AV 14.} 
AV~14 was first classified as ``O5 + neb" by Ardeberg \& Maurice (1977),
and reclassified as ``O3-4~V + neb" by Garmany, Conti, \& Massey (1987).
Its early spectral type resulted in its inclusion in our program.  The
``+ neb" designation often came about due to
photographic spectral classification
of early-type stars in the Magellanic Clouds, as local sky subtraction was
not possible as it is with CCD spectrometers. Here we reclassify the star as O5~V.  The visual appearance of the spectrum (Fig.~\ref{fig:AV14}a)
is consistent with the
measured ratio of the equivalent widths, $\log W'=-0.57$.
The absolute visual magnitude of this star, $M_V=-5.8$, would suggest it is intermediate between luminosity class ``V" and `III", according to
Conti (1988); however, by comparison to any Galactic standards, the luminosity
class is clearly ``V", as He~II $\lambda 4686$ is strongly in absorption.
The spectrum is a very good match to that of 
Walborn \& Fitzpatrick's (1990) spectrum of HD~46150, although there is no
hint of N III $\lambda\lambda 4634,42$ emission in our spectrum.

We obtained good fits after just a few models (Fig.~\ref{fig:AV14}b).  
In this case, the He~I $\lambda 4387$ line strength
would suggest a slightly lower temperature (42000$^\circ$K) than the
one we adopt here.  However, this lower temperature produces He~II lines
which are weaker than observed. The He~I $\lambda 4471$ line shows
good agreement with the fits based upon He~II.  The mass-loss rate is 
low, and therefore not very well determined, with adequate fits obtained
with values $\dot{M}\le 3 \times 10^{-7} M_\odot$ yr$^{-1}$.

{\it AV 26.} 
AV~26 was classified as an O7 III by Garmany et
al.\ (1987).  Here we classify it as just slightly earlier, O6, based 
upon
the visual impression of the spectrum (Fig.~\ref{fig:AV26}a)
We also measure $\log W'=-0.22$, 
consistent with the O6 spectral type. As for
the luminosity class, the star's absolute visual luminosity
($M_V=-7.0$) requires it to be a bright supergiant.  A Galactic O6
star of this luminosity, though, would have very strong N~III $\lambda\lambda
4634, 42$ and He~II $\lambda 4686$ emission.  Instead, we see weak
N~III emission and slightly weakened He~II $\lambda 4686$ {\it
absorption}.  Nor would we expect to see N~IV $\lambda 4058$ emission
in an O6 star of any luminosity type.
Are these peculiarities due to the low metallicity characteristic of the
SMC, or are we seeing a composite spectrum?  The presence of N~IV emission
in Galactic O-type 
stars is invariably either coupled to strong He II $\lambda 4686$
emission (in an O3-4 I), which we don't see, or to N~V absorption
(in an O3 III), which we also don't see.  (N~IV $\lambda 4058$ in dwarfs
is weaker than what we detect in our spectrum, -70m\AA.)
Given this, we tentatively
rule out the composite explanation\footnote{We are grateful to Nolan
Walborn for correspondence on this subject, although our final conclusion
differed from his, primarily because at our high SNR (350 per 1.2\AA\
resolution element) we can place a very stringent upper limit on the
presence of any N~V absorption.}.  We believe the spectral features
we see in AV~26 are just normal for an O6 I star in the low metallicity
environment of the SMC. 
We classify this spectrum as O6 I(f).  The following star (AV~75) provides
an additional example.
Unless the star {\it is} a composite spectral type, then the behavior
of the N~III and N~IV selective emission lines underscores the dangers
of interpreting these line ratios as an effective temperature indicator,
as was recently done by Walborn et al.\ (2002a, 2004).

We judge the fits of the model to the observed spectrum good. These are
shown in Fig.~\ref{fig:AV26}(b).  The surface gravity is $\log g=3.5$, 
consistent with the supergiant designation.
AV~26 is one of the two stars for which we also have near-UV data for
the He~I $\lambda 3187$ and He~II $\lambda 3203$ lines. We will
introduce these powerful diagnostic lines in Section~\ref{Sec-nearUV}.

{\it AV~75.}
AV~75 was classified as an O5~III(f) star by Garmany et al.\ (1987), and
as O5~III(f+) by Walborn et al.~(2000), who make reference 
to its ``entirely normal hot O giant spectrum" in both the optical and UV.
Our spectrum (Fig.~\ref{fig:AV75}a)
agrees well with the ``O5" designation, and we measure $\log W'=-0.38$.
However, at $M_V=-6.9$ the star must be considered a supergiant but with
He~II $\lambda 4686$ in absorption rather than emission due to the smaller
stellar wind that comes from lower metallicity.  
The ``+" designation signifies that Si~IV $\lambda 4116$ is in emission, 
a result
which we also confirm.  Like Walborn et al.\ (2000), we note that the
Si~IV $\lambda 4089$ line is very weakly present in absorption, 
a result which we find somewhat curious despite
 Walborn et al.\ (2000)'s explanation.  We classify the star as
O5.5~I(f+). Good agreement is found is found with the model fits
(Fig.~\ref{fig:AV75}b).

{\it AV~207.} Our spectrum (Fig.~\ref{fig:AV207}a) of AV~207 shows
 He~I $\lambda 4471$ is just slightly stronger than 
He~II $\lambda 4542$ in this star, making it of spectral type O7.5.
We measure $\log W'=+0.03$.
N~III is very weakly in emission, with He~II $\lambda 4686$ 
absorption strong, and
so we designate the star as ``((f))".  The absolute luminosity ($M_V=-5$)
is consistent with the star being of luminosity class ``V" (i.e.,
Conti 1988).   Previously the star was classified as O7~V by 
Crampton \& Greasley (1982), in good agreement with our determination of
O7.5 V((f)).  A good fit (Fig.~\ref{fig:AV207}b)
was found for all of the lines.  The mass-loss rate for this star is low,
with values $\le 3 \times 10^{-7} M_\odot$ yr$^{-1}$ yielding good fits.

{\it AV 296.} At first glance (Fig.~\ref{fig:AV296}a), 
AV 296 appears to simply be
a broader-lined version of AV~207.  We classify it it similarly
as O7.5 V((f)) visually, and measure $\log W'=+0.05$, consistent with that
spectral subtype.  The absolute visual magnitude is very similar ($M_V=-5.1$)
to that of AV~207.   The star was previously classified as ``O5 V:" by
Garmany et al.\ (1987), with the ``:" denoting an uncertain type due to 
nebular contamination.  

We obtained a barely adequate fit (Fig.~\ref{fig:AV296}b)
to the spectrum.  The required surface
gravity $\log g$ is low, more characteristic of a supergiant than a giant or
dwarf.
Possibly this is an effect of the rapid rotation lowering the effective
surface gravity, but we suspect that this star may be a spectroscopic
binary with not-quite-resolved double lines at the time of the exposure.
This view is further supported by the measured radial velocity of the star,
$\sim 250$ km s$^{-1}$, which is quite high given the 158~km s$^{-1}$
systemic velocity of the SMC (see Fig.~1 of Massey \& Olsen 2003).
A radial velocity program is probably warranted, but for now we include the derived values in Table~4 but note this uncertainty.

{\it AV 372.} We originally classified this star as an O9.5~I 
(Fig.~\ref{fig:AV372}). Walborn et al.\ (2002b) arrived at a similar
type (``O9~Iabw").
The supergiant status is suggested
not only since $M_V=-6.8$ but also due to the strength of Si~IV absorption.
However, we were unable to find any simultaneous good agreement of the
strengths of both He~I and He~II.  In addition, the H$\alpha$ profile
appears to be P Cygni, and no amount of tweaking of the mass-loss rates
and $\beta$ (from 0.5 to 2.5) produced an acceptable fit at the temperatures
indicated by the He~I to He~II ratios.  The Balmer line profiles were 
suggested of very low surface gravity ($\log g=3.2$), and at first we
thought that we had encountered a problem with the models.  However,
further inspection of the optical spectrum revealed that the He~I lines
were significantly {\it broader} than the He~II lines.  We measure a
$v \sin i$ of 110 km s$^{-1}$ for He~II but require a 
$v \sin i$ of 200 km s$^{-1}$ for He~I.  The centers of He~I lines are also
shifted by -20 km s$^{-1}$ with respect to those of He~II.
The star is
likely a spectroscopic binary, with two stars contributing to He~I and
one star dominating the He~II spectrum.  

{\it AV 377.}
The spectrum (Fig.~\ref{fig:AV377}a)
appears to be that of an O5~V((f)), with slightly
anomalously strong N~III $\lambda\lambda 4634,42$ emission.  Even visually, however, we
see that the He~II lines are rather strong compared to hydrogen,
in comparison to the Walborn \& Fitzpatrick (1990) atlas.  We measure
a value of $\log W'=-0.55$, consistent with the O5 spectral subtype.
The absolute magnitude, $M_V=-4.9$, is consistent with the luminosity
class ``V" designation.  Previously, the star was classified as O6~V by
Garmany et al.\ (1987). 

Fitting this star required increasing the He/H number ratio from the
canonical value 0.10 to a considerably higher value: 0.35.  When we did this
we obtained simultaneously good fits to the He~I, He~II, and Balmer lines
(Fig.~\ref{fig:AV377}b).  
The only exception was He~II $\lambda 4542$, for which the model
line was weaker than the observed line.  However, 
the fits at He~II $\lambda$4200 and He~II $\lambda 4686$ were good.  
Increasing the temperature slightly
does not improve the fit to He~II $\lambda 4542$ and makes the model He~I
spectra too weak.  Similarly, either a slighter lower value (0.3) or 
higher value (0.4) for He/H produced He lines that were too weak or too
strong, respectively.

It has become increasingly recognized that rotation can play an important
role in enriching the surface material even during the main-sequence stage
(Maeder \& Meynet 2000, Walborn et al.\ 2004).  
The enriched He abundance 
is likely consistent with the stronger-than-expected 
presence of N~III $\lambda\lambda 4634,42$ emission.
The low mass-loss rate we find for this star ($\dot{M}\sim 10^{-7} M_\odot$ yr$^{-1}$, with only an upper 
limit $\dot{M}\le 3 \times 10^{-7}M_\odot$ yr$^{-1}$ firmly 
established) is surprising given its
spectral type, but is consistent with the UV spectrum, for which the
stellar wind lines were quite weak.

{\it AV 378.}
We classify this star as spectral type O9.5~III (Fig.~\ref{fig:AV378}a)
with guidance from the referee Nolan Walborn, who argues that He~II
absorption is consistent with a giant classification. We initially
preferred a supergiant designation.  The spectral subtype is consistent
with $\log W'=+0.47$. We find $M_V=-5.5$, which is halfway between what
we expect for a giant and supergiant.  The star has been previously
classified as O8~V by Garmany et al.\ (1987).

The surface gravity of this star is low ($\log g=3.25$),
which is much more consistent for a supergiant than that of a giant.
We quickly found that we needed to slightly increase the He/H ratio to
obtain an adequate fit for both the He~I and He~II lines
(Fig.~\ref{fig:AV378}b).  For this star, the He~I $\lambda 4471$ and
He~II $\lambda 4387$ lines were inconsistent, in the same sense as
described by Repolust et al.\ (2004), i.e., the model spectrum of He~I
$\lambda 4471$ is too weak at lower temperatures in giants and
supergiants.  Accordingly we have relied upon the He~I $\lambda 4387$
and to a lesser extent on the He~I $\lambda 4922$ line (not
illustrated) in determining the fit.

This star did not have any measurable stellar wind lines, and so we
simply adopted a value of $v_\infty=2000$ km s$^{-1}$ in computing its
models.  We lack data at H$\alpha$, but began with the initial
assumption that the mass-loss rate was quite low ($10^{-7} M_\odot$
yr$^{-1}$).  Examination of the He~II $\lambda 4686$ profile then
suggests that $\dot{M}$ may be even lower.
The temperature, surface gravity, and
He/H ratio are all very well determined in this regime.

{{\it AV 396.}
We classify the spectrum of AV~396 as B0~III (Fig.~\ref{fig:AV396}a).  The
spectral type is clearly later than that of AV~378 (O9.5~III).  Classically,
the dividing line between O9.5 and B0 was the presence or lack of He~II
in the spectrum (Jaschek \& Jaschek 1990), but higher signal-to-noise
spectra 
now results in stars of B0 type having detectable He~II $\lambda 4200$ and
He~II $\lambda 4542$ as well as the strong (luminosity-dependent)
He~II $\lambda 4686$ line.  If $\upsilon$~Ori (HD~36512) is considered
a B0~V, as all authors have done since Johnson \& Morgan (1953), then
we would conclude that the spectral type of AV~396 is also B0.  A later
type (such as B0.2, as introduced for $\tau$ Sco by Walborn 1971a), can be
ruled out based upon the fact that Si~IV $\lambda 4089$ is strong but
Si~III $\lambda 4552$ is all but non-existent, while these lines are of
comparable strengths in $\tau$ Sco\footnote{Here we forced to
eschew the intermediate
class O9.7 introduced by Walborn (1971a) and used by Walborn \& Fitzpatrick
(1990).  First, this intermediate class is defined only for supergiants,
and secondly, it is based upon the ratio of the He~II $\lambda 4542$ to
Si~III $\lambda 4552$ lines.  The strength of the latter is not only
gravity-dependent, but the strength relative to He~II will also depend
heavily on the metal-content of the star.}.  Since we can determine the
physical parameters of this star based upon the He~I to He~II ratio, we
consider the ``B0" spectral type an honorary member of the O-type class.
A comparison of the properties derived in this way will be made to
what we obtain using the Si~IV to Si~III lines in a subsequent paper.
The absolute magnitude of the star,
$M_V=-5.2$, is consistent with it being a giant, although (as expected)
the Si~IV lines are weaker with respect to He~I what we would find
in a Galactic giant. (Compare Fig.~\ref{fig:AV396}a with the spectrum
of HD~48434 shown in Fig.~17 of Walborn \& Fitzpatrick 1990).
Previously the star was classified as O9~V by
Garmany et al.\ (1987), doubtless
due to the (weak) presence of He~II.

We did not detect any measurable stellar wind lines in the UV, and we lack
an H$\alpha$ spectrum of the star. We again assume a minimal mass-loss
rate and a terminal velocity $v_\infty=2000$ km s$^{-1}$ in making the fit.
We judge the fits shown in Fig.~\ref{fig:AV396}b excellent, with the
values of $T_{\rm eff}$, $\log g$, and He/H well determined. (We did
have to increase He/H to 0.15 to obtain a good fit.)  The model
He~I $\lambda 4471$ line is again weaker than expected compared to
the other He~I lines, primarily He~I $\lambda 4387$.

{\it AV~451.} The spectrum of this star 
(Fig.~\ref{fig:AV451}) is clearly earlier
than that of AV~396 (Fig.~\ref{fig:AV396}), and very similar to that
of AV~378 (Fig.~\ref{fig:AV378}a), 
except that the luminosity indicator Si~IV to
He~I is much weaker in AV~451.  We classify this spectrum as
O9.5~III, consistent with $M_V=-5.2$  Previously this star was called an
O9~V by Garmany et al.\ (1987). 

Again, we were not able to discern any stellar wind lines in our UV
spectrum of the star, and we lack an observation of the H$\alpha$ profile.
We assume a minimal mass-loss rate ($10^{-7} M_\odot$ yr$^{-1}$) and
terminal velocity $v_\infty=2000$ km s$^{-1}$ in making the fit.

Our first attempt at modeling the spectrum of this star, however, revealed
that it is a likely double-lined spectroscopic binary: the He~I lines are
much broader ($v \sin i = 200$ km s$^{-1}$) than the He~II lines 
($v \sin i = 120$ km s$^{-1}$), suggesting that this system consists of
a mid-O and late-O pair of stars.  No satisfactory combination of surface
gravity and rotational velocities could match the Balmer lines. We add this
to our list of stars that deserve radial velocity monitoring.

{\it AV~469.}
The spectrum of this star is readily classified as O8.5 I(f), with the only
spectral peculiarity being that He~II $\lambda 4686$ is 
in absorption,
while in a Galactic O8 I star of comparable $M_V$ we would expect it to
be mostly filled in by emission (e.g., HD 17603; see Conti \& Alschuler 1971).
Otherwise, the spectrum is very similar to that
shown by Walborn \& Fitzpatrick (1990) for HD~151804 (O8 Iaf), 
including
the strong 
N~III absorption features at 4097\AA\ and 4511-15\AA (Fig.~\ref{fig:AV469}a).
We measure a value of $\log W'=+0.24$, which suggests the intermediate
(O8.5 rather than O8) type.
The absolute visual magnitude of the star, $M_V=-6.2$, is consistent with
its supergiant designation.  Previously, this star was classified as O8~II
by Garmany et al.\ (1987), and as O8.5~II((f)) by Walborn et al.\ (2002b).

Despite running 28 models for this star, we were left unsatisfied with the
final fit (Fig.~\ref{fig:AV469}b).  
In particular, the H$\alpha$ line has a small emission bump
which we were unable to reproduce despite our exploration of parameter
space (both $\beta$ and $\dot{M}$).  In addition, the velocity of
the synthetic He~II $\lambda 4686$ line is clearly shifted to the blue
relative to the observed line.  Nevertheless, values of $T_{\rm eff}$ and
$\log g$ seem well determined.  He~I $\lambda 4471$ was much weaker in
the synthetic spectrum than in the observed spectrum, consistent with
our experience that this occurs at low surface gravities and relatively
``cool" temperatures.

\subsubsection{The non-R136 LMC Stars}

{\it LH64-16.}\footnote{Note that this star has sometimes been referred
to as ``W16-8", as it is also star 8 in Field 16 of Westerlund (1961).
The LH64-16 designation is from Lucke (1972).  A hybrid version of the
name, ``LH64W8", was unwisely used as the designation for the {\it HST}
observations.}. Massey, Waterhouse, \& DeGioia-Eastwood (2000) classified LH64-16 as ``O3 III:(f*), where the
``*" notation denotes that N~IV$\lambda 4058$ emission is stronger
than N~III $\lambda\lambda 4634, 42$ emission, which is generally a characteristic
of O3 supergiants (Walborn \& Fitzpatrick 1990). 
We illustrate the
spectrum in Fig.~\ref{fig:LH6416}a.  Although ``classically" the O3
spectral type is one which lacks He~I absorption (Walborn 1971b, Conti 1988), 
high SNR data can reveal weak ($W=\sim 75$ to 250 m\AA) He~I $\lambda 4471$
(Kudritzki 1980, Simon et al.\ 1983).  We do detect He~I $\lambda 4471$ 
very weakly in our spectrum; the measured strength is about 100m\AA.

The absolute visual magnitude of LH64-16 is only $-5.2$, which would suggest 
it is a dwarf.  However, there is a general problem with this argument when
applied to a degenerate spectral class.  A 55,000$^\circ$K O star (were such
an object to exist) and a 48,000$^\circ$K O star would both be 
classified as ``O3" (as they would lack significant He~I).  If they had the
same bolometric luminosity, then the hotter star would be visually fainter.
(Since stars evolve at fairly constant $M_{\rm bol}$ this situation could
apply simply to the same star at two slightly different ages; the fainter,
hotter star would be the younger.)
Thus if LH64-16 were a particularly hot O3 star, then it could well be of
luminosity class III.  From a morphological point of view, there are conflicting
data on the luminosity criteria: the strength of N~V $\lambda\lambda 4603, 19$
absorption, and N~IV $\lambda 4058$ emission would argue this is a supergiant,
while the presence of He~II $\lambda 4686$ absorption would argue that it is
a giant or a dwarf.

Walborn et al.\ (2002a) used our spectrum of the star to help define a
new spectral class: that of O2.  This was based on the relative strength
of N~IV $\lambda 4058$ and N~III $\lambda\lambda 4634,42$ emission and N~V
$\lambda 4603, 19$ absorption.  The implications are, of course, that O2
is a ``hotter" spectral type than O3, but the interpretation is complicated
by the fact that the N lines are luminosity (gravity?) dependent. 
Walborn \& Fitzpatrick (1990) nicely illustrate this in their Fig.~8.  There is
not a good theoretical underpinning of this new spectral type, as yet.  
Walborn et al.\ (2002a) consider LH64-16 to be of an O2~III(f*) star.

The situation has been further complicated by the discovery that LH64-16
is nitrogen-enriched.  Walborn et al.\ (2004) produce yet another new
spectral designation, namely ON2~III(f*) to describe the spectrum.  Their
modeling of our spectrum suggests a somewhat
enhanced He/H number ratio (0.25) and extremely high
temperature (55,000$^\circ$K), where model fits were obtained by CMFGEN
using the near-UV and optical spectrum.  (Values of surface gravity were
adopted, so the issue of luminosity class remains unresolved.) 
The very high temperature is unprecedented for a model fit with 
fully-blanketed models.

Our own modeling of this star is shown in Fig.~\ref{fig:LH6416}b, where
we have fit only the optical lines.  We
found that we needed a high He/H ratio if we were to get the model He~II
lines as strong as what is observed, and we were forced to
increase the He/H ratio to 1.0 to obtain a satisfactory fit.
Once we had increased the He/H ratio, though, we needed a very 
high effective temperature in order to make the
He~I lines as weak as that observed.  
Our value for the surface gravity
is well constrained (as usual) by the wings of the Balmer line profiles
to $\log g=3.9$, suggesting that it may be a dwarf.  The He~II $\lambda$
4686 profile is very sensitive to the mass-loss rate, and the value we
derive from H$\alpha$ gives a reasonably good fit.  
Given that
Walborn et al.\ (2004)
were also able to model the atmospheric
abundances, including the variations of CNO 
with CMFGEN, their parameters may be better determined than
ours, although the source of the disagreement in the He/H ratio (our 1.0 vs.\
their 0.25) is hard to
understand, especially given the fact that they were forced to a similarly
high temperature in order to fit the star. 
They present this star as a possible example of ``homogeneous
evolution"---that somehow, possibly due to rapid rotation, the star
has evolved chemically in such a way that the surface composition is
similar to that of the core.  The nitrogen enhancement found by 
Walborn et al.\ (2004) is a factor of 7 over the presumed starting
value. This much nitrogen is consistent with any 
He/H ratio from 0.25 to 2.0, as it is simply the
CNO-burning equilibrium ratio (see Massey 2003
and references therein.)

Our modeling generally
supports their results, although not necessarily their interpretation.
We find a ``spectroscopic mass" ($M \sim g/g_\odot \times R^2$)
of $26 M_\odot$.  A similar value is necessitated by the model of Walborn
et al.\ (2004).  Yet, the ``evolutionary mass" found by Walborn et al.\ (2002a)
is much higher, 72$\cal M_\odot$, based upon the (non-rotation) evolutionary
models of Schaerer et al. (1993).  Although for some time there were hints
of a mass discrepancy between the spectroscopic and evolutionary masses,
improvements in the models (both atmospheric and evolutionary) have 
largely eliminated this problem (Repolust et al.\ 2004 and references therein).
We will revisit this topic in the second paper in this series.  However, here
we offer the suggestion that the relatively low mass inferred by the
atmosphere modeling (both ours and that of Walborn et al.\ 2004) is
connected to the high chemical abundances found at the surface.  Possibly
this star is the result of binary evolution, or some other peculiarity.
We are indebted to Nolan Walborn for calling this discrepancy to our
attention.

{\it LH81:W28-5.}
The spectrum of LH81:W28-5  (Fig.~\ref{fig:LH81w5}a)
is readily classified as O4~V((f+)),
with $\log W'=-0.68$.  The presence of weak N~V
4603,19 absorption is consistent with this classification.
The ``+" designation denotes the presence
of Si~IV $\lambda \lambda 4089, 4116$ emission.  Si~IV emission
is not seen in
the example of an O4 V((f)) star shown by Walborn \& Fitzpatrick (1990),
but our data are of higher SNR, and indeed the Si IV emission features are
seen in the Galactic star HDE 303308 described as O4 V((f+)) by Walborn
et al.\ (2002a). The faint absolute visual magnitude
($M_V=-5.0$) is consistent with the strong He~II $\lambda 4686$ absorption
feature in determining the ``V" luminosity class.  Walborn et al.\ (2002a) 
cites this star as a representative of the O4V((f+)) class.

Our fit of the parameters of this star was straight-forward, other than
the fact the He/H ratio had to be increased slightly in order to produce He~II
lines as strong as those observed.  We judged the agreement between the
models and the observations very good, and show the comparison in 
Fig.~\ref{fig:LH81w5}b.

{\it LH101:W3-24}\footnote{This star was also cataloged as ST5-27 by
Testor \& Niemela (1998).  It was observed with {\it HST} under the
{\it nomme de plume} ``LH101W24".}.
Our ground-based spectrum of LH101:W3-24 (Fig.~\ref{fig:lh101w24}a) is strongly
contaminated with nebular emission lines at the Balmer lines, despite
our best efforts to avoid them by narrowing the slit and attempting various
regions for sky subtraction.  We were fortunate to be able to supplement
our ground-based data with exposures both in the near-UV and at H$\alpha$ with {\it HST}, allowing us to use a very narrow (0.2 arcsec) slit.  (We will
discuss the near-UV spectrum below in Sec.~\ref{Sec-nearUV}.)  Previously,
the spectrum had been classified as O4~V by Testor \& Niemela (1998), 
presumably because He~I $\lambda 4471$ is very weakly present.
We measure an equivalent width of 120m\AA\ for this line, placing it
in the same regime as other O3 stars.  We thus retain the O3~V((f))
designation of Massey et al.\ (2000).

Fortunately, the strong nebular lines did not interfere with the fitting:
our uncontaminated {\it HST} H$\alpha$ spectrum was used to determine
the mass-loss rate, while the ground-based Balmer line observations served
adequately for the determination of $\log g$, which after all is based
upon the fits to the Balmer wings.

We obtained adequate fits, although we did have to slightly increase the
He/H ratio from 0.10 to 0.15 to make the He~II lines sufficiently strong.
There was some disagreement between He~II $\lambda 4200$ and He~II $\lambda 4542$, and we arrived at a compromise. The spectra are somewhat more noisy
than most due to the need for a narrow extraction aperture to reduce
the effects of the nebular emission. In addition the He~II $\lambda 4200$ 
line is compromised somewhat due to a bad column which sat on the red wing
of the profile.  Nevertheless, the overall agreement (Fig.~\ref{fig:lh101w24}b)
is good.

\subsubsection{The R136 Stars}

The spectroscopic study of R136 by Massey \& Hunter (1998) identified more
O3 stars than had been previously known in total elsewhere.  The FOS spectra
that were used for that study were of relatively poor SNR, and suffered from
intermittent behavior of some of the diodes, limiting the ability to 
flat-field well.  As described in Sect.~\ref{Sec-data}, we obtained higher
quality data with STIS in a follow-up study designed to allow the modeling
we now describe.  However, the STIS data covered only a very limited 
wavelength range (from H$\gamma$ through He~II $\lambda 4542$).  In what
follows we make use of the older FOS data in describing the overall
spectrum (and determining the spectral type), but restrict our modeling
only to the STIS data.  The fits to the FOS's He~II $\lambda 4686$ 
observations are shown, however, for comparison.  In presenting the 
blue-optical spectra, we have spliced in the better STIS data in the
wavelength region 4310\AA\ to 4590\AA. 

{\it R136-020.}  This star was classified as O3 If*/WN6-A, with the
``slash" designation a tribute to the very strong emission features
at N~IV $\lambda 4058$ and He~II $\lambda 4686$, with strengths and
widths comparable to those produced in the stellar wind of a Wolf-Rayet
star.  A comparison of Fig.~\ref{fig:r136020}a with that of
the O3 If* star HD 93129A shown in Walborn \& Fitzpatrick (1990) shows
that this is simply a more extreme example, and classically one could
 drop the
``slash" part of the designation and simply call the star an O3 If*. 
The star was left off the list of O3 stars by Walborn et al.\ (2002a)
due to its ``slash" description. However, by their
classification criteria it would be called an O2~If*, given the
strength of N~IV $\lambda 4058$ emission and the lack of any He~I absorption
or N~III emission.
The absolute magnitude $M_V=-6$ is consistent with the star's supergiant 
designation.

Despite the high mass-loss rate implied by both H$\alpha$ and 
He~II $\lambda 4686$, the fitting of this star was straight-forward,
and good matches were achieved. 
A slightly elevated He/H ratio was needed in order to make the He~II
$\lambda 4542$ line as strong as observed.  He~I $\lambda4471$ is 
not detected even in our high SNR spectrum, and thus the effective
temperature (and hence bolometric luminosity) given in Table~4 must
be considered a lower limit.

{\it R136-024.} This star was classified as O3 III(f*) by Massey \& Hunter
(1998); its spectrum is shown in Fig.~\ref{fig:r136024}.  We were unable
to obtain a good fit, and the broadness of the spectral lines suggests
it is an incipiently resolved double-lined binary.  Indeed, after we began
this project, Massey et al.\ (2002) found light variations that were
typical of eclipses. Given the weakness of He~I in the spectrum, this system
likely consists of two O3~V's, and thus is a good candidate for radial
velocity studies with {\it HST}.  Such a project is being proposed.

{\it R136-036.}  This star was classified as O3 If* by Massey \& Hunter
(1998); its spectrum is shown in Fig.~\ref{fig:r136036}a.  The spectrum shows
the same luminosity features of the O3 If* star R136-020, i.e., strong
NIV $\lambda 4058$ and He~II $\lambda 4686$ emission, and N~V $\lambda 4603, 19$
absorption.  Walborn et al.\ (2002a) refer to this as an ``O2-O3 If*"
star under the designation ``MH36".  The ambiguity in the spectral type was
due to the noisy region around He~I $\lambda 4471$ in the FOS spectrum;
with our higher SNR STIS spectrum, they would undoubtedly have considered
this an O2~If* star given the lack of He~I and the fact that the N~IV emission
is so much stronger than any possible N~III $\lambda\lambda 4634,42$ emission.
We adopt this designation here, despite our ambivalence about the use
of selective emission features to extend the spectral classification.

Our fits of H$\gamma$ and He~II $\lambda 4542$ are quite good; the effective
temperature given in Table~4 is the lowest for which we obtain a sufficiently
weak He~I $\lambda 4471$ line, and thus must again be considered a lower
limit.  Although we investigated a broad area of parameter space ($\beta$
and $\dot{M}$) we did not obtain good fits to the stellar-wind sensitive
lines H$\alpha$ and He~II $\lambda 4686$.  Possibly this is a consequence
of the lack of wind clumping in our models, as H$\alpha$ is strongly in
emission. 

{\it R136-040.}  This star was classified as O3~V by Massey \& Hunter (1998);
its spectrum is shown in Fig.~\ref{fig:r136040}a.  Since neither
NIV~$\lambda 4058$ nor N~III $\lambda\lambda 4634,42$ emission 
is visible, we would be hard-pressed to refine the classification (O2-O3.5)
using the
scheme of Walborn et al.\ (2002a).
There is no He~I $\lambda$ 4471 detected in this spectrum.

In order to produce He~II $\lambda 4542$ as strong as what is observed, we
must increase the He/H ratio to 0.2.  That then requires a very high
$T_{\rm eff}$ to match the stringent limit on the He~I$\lambda 4471$ line.
Again this effective temperature should be treated as a lower limit, since
no He~I$\lambda$ 4471 is actually detected, despite the excellent SNR
(150 per 2-pixel resolution element)
and resolution of our data.

{\it R136-047.}  This star was classified as O3~III(f*) by Massey \& Hunter
(1998); its spectrum is shown in Fig.~\ref{fig:r136047}a.  Despite the 
higher SNR data at He~I $\lambda$ 4471, no trace of this line could be found.
The giant luminosity class is due to the presence of N~IV $\lambda 4058$
emission.  By the criteria of Walborn et al.\ (2002a), this would have to be
called an O2~III(f*) star.  We obtained good matches of the model spectrum
to the observed spectrum for this star (Fig.~\ref{fig:r136047}b), 
but the fit required a very high
effective temperature ($T_{\rm eff}$=51,000$^\circ$K) in order to make the
model He~I $\lambda 4471$ line sufficiently weak.
Even so, we must consider this a lower limit, since {\it no} He~I $\lambda 4471$
is detected in our spectrum.  
Although we explored much of parameter space
(30 models), we never found a combination of 
$\beta$ and $\dot{M}$ that produced
as good a fit to the H$\alpha$ profile as we would have liked.

{\it R136-055.}
This star was classified as O3~V by Massey \& Hunter
(1998); its spectrum is shown in Fig.~\ref{fig:r136055}a. Weak He~I $\lambda 4471$ is present, allowing an accurate effective temperature to be determined.
We were able to fit the spectrum of this star very straight-forwardly
(Fig.~\ref{fig:r136055}b), deriving a high effective temperature and
a surface gravity consistent with its being a dwarf.

\section{Introducing the He~I $\lambda 3187$ and He~II $\lambda 3203$ 
Diagnostic Lines}
\label{Sec-nearUV}

Morrison (1975) described coud\'{e} observations of the He~I $\lambda 3187$
and He~II $\lambda 3203$ lines in O and early B-type stars
made from the high altitude of Mauna Kea.  She motivated this discussion
by noting that the He~II $\lambda 3203$ absorption line ($n=3$ to 5) 
provided a unique opportunity to test stellar 
atmosphere models, as it was the only accessible He~II
line that did not involve
$n=4$ (i.e., all of the Pickering lines arise from $n=4$; e.g.,
He~II $\lambda 4200$, $n=4$ to 11; $\lambda 4542$, $n=4$ to 9).
Given the presence of emission at He~II $\lambda 4686$ ($n=4$ to 3) in
many O-type supergiants,
Morrison (1975) argued that the $n=4$ level is overpopulated, 
at least in some parts of
the stellar atmosphere for some stars, although today we would instead
say that the emission is simply a pure wind effect, due to the large
contributing volume, and has nothing to do with overpopulation.
Morrison (1975) found that the He~II $\lambda 3203$
line was much weaker than predicted by the plane-parallel non-LTE models of
Auer \& Mihalas (1972) for the stars in which He~II $\lambda 4686$ is in
emission, although reasonable agreement was found in the other cases.
The He~I $\lambda 3187$ line falls in the same spectral region but there
were no theoretical predictions for the strength of this line at the time.

We are unaware of any follow-up of this interesting work, doubtless due to
the difficulty of observing this wavelength region from the ground.
Despite the numerous observations of OB stars with {\it IUE} (in operation
from 1978 to 1996), no new studies of this line were made, probably
due to the proximity of the He~II $\lambda 3203$ line 
to the long-wavelength cut-off.  However, this wavelength region
is easily observed with {\it HST}.  We have long been intrigued by the
Morrison (1975) paper, and when it became apparent that there was enough
time remaining in the visibility period for a few of our stars to allow
observations in this region after our primary observations were complete,
we availed ourselves of this opportunity (see Sec.~\ref{Sec-optical}).  
We did not use these lines 
to determine the model fits previously described; instead, we simply use the
near-UV observations here to set the stage for future work,
by comparing with our model fits.

We have data in the near-UV for two of the stars in this paper, AV~26
and LH101:W3-24.  The first of these is an O6 I(f) supergiant, and the
latter an O3 V((f)) dwarf.  Neither has He~II $\lambda 4686$ in emission,
however.  The spectrum of AV~26 has a good SNR (100 per 2-pixel spectral
resolution element), while the spectrum of LH101:W3-24 is much noisier
(SNR=11 per 2-pixel spectral resolution element).

We show the spectra and the model predictions
in Fig.~\ref{fig:3200}.  First, we
note that neither star has measurable He~I $\lambda 3187$ in its
spectrum.  The adopted model for AV~26 predicts a slightly stronger
He~I $\lambda 3187$ than observed, while the agreement for LH101:W3-24
is as good as the signal-to-noise allows.  Second, the He~II $\lambda 3203$
line is very well matched for AV~26.  The synthetic He~II $\lambda 3203$
spectrum may not be strong enough in LH101:W3-24; it is hard to tell,
given the poor SNR.  Further comparisons are planned in the next paper
in this series.

\section{Conclusions and Summary}

We have attempted to model the spectra of 20 O-type stars in the 
Magellanic Clouds, succeeding in obtaining adequate or good
fits in 17 cases.
We suspect that the other three stars are binaries, and expect that
a few more in our sample are in reality composite spectra, given
the statistical expectations that a third of (Galactic)
O-type stars are spectroscopic binaries.  The physical parameters of the
stars which were successfully modeled are well determined, and should
permit us to refine the effective temperature scale.

In describing the spectral features in our sample, we were struck by
the fact that many of our SMC stars appear to be more luminous than
their ``f" characteristics (N~III $\lambda 4634, 42$ and He~II $\lambda
4686$) would indicate.  Although one cannot rule out a binary
explanation, we did not see such discrepancies amoung our LMC sample,
suggesting that the effects of lower metallicity may be responsible.
N~III emission is a complex NLTE effect dependent primarily on temperature,
while He~II $\lambda 4686$ is formed in the stellar wind, and its formation
will be affected by the temperature, wind density, and amount of EUV flux,
all of which will be affected by the metallicity.  Thus one would not
expect {\it a priori} that these ``f" characteristics would exhibit the 
{\it same} behavior at low metallicity as they do in Galactic objects.  
We have identified here everal examples where the He~II emission is weaker 
than one might expect given the star's $M_V$, while the referee Nolan 
Walborn has kindly reminded us of two additional examples in the SMC, Sk~80 
and AV~83, where the He~II $\lambda 4686$ emission is possibly {\it
stronger} than in many Galactic counterparts.  More modeling of both
Galactic and SMC stars are needed to understand this issue.

In the second paper of this series we will roughly double the sample of
Magellanic Cloud stars that have been analyzed in this manner.  However,
it is tempting to compare the effective temperatures we have derived here
with those recently found for Galactic stars 
by Repolust et al.\ (2004).  We show this
comparison in Fig.~\ref{fig:results}, where we have separated the effective
temperatures by luminosity class.  For comparison, we also show the
effective temperature scales determined by Conti (1988) (dashed line)
and Vacca et al.\ (1996).

The data for the giants are too sparse as yet to draw any conclusions.
However, for both the dwarfs and the supergiants we see that the Magellanic
Cloud stars have {\it significantly} higher temperatures than their
Galactic counterparts in the range from the earliest types through mid-O.
By the late O-types, there is less of a difference.
The data are also too sparse yet to draw firm conclusions about differences in
the effective temperature scales for O stars for the SMC vs.\ LMC.
In the sample we have analyzed so-far we have mainly very early O-type
stars in the LMC, and somewhat later types in the SMC.  We will be able
to address this more fully in the second paper in this series, where we
complete our Magellanic Cloud sample.

Nevertheless, this result is quite intriguing.  As described in Sec.~\ref{Sec-hist}, we
would expect the lower mass-loss rates (due to the lower
metallicities) in the Clouds to result in
a higher effective temperature in comparison to a star of the same 
spectral subtype in the Milky Way: first, there will be less wind emission
affecting the He~I.  In addition, the smaller metallicity will lead to 
reduced {\it wind-blanketing} and {\it line-blanketing}.  Together,
these combine to 
result in effective temperatures which are 
{\it $\sim3000-4000^\circ$K} (10\%) greater at O5~V for the Magellanic Cloud
stars in our sample.

We note with some irony that despite the substantial improvements in the
stellar models over the years, the older effective temperature scales
of Conti (1973) and Conti (1988), based primarily on the original 
non-LTE models of Auer \& Mihalas (1972), have held up remarkably well for
the Galactic stars.  Work over the next few years will result in an improved
scale that will take metallicity into account, but for now the Conti (1988)
scale is to be preferred over more recent editions (e.g., Vacca et al.\ 1996).

Three recent studies have analyzed a limited sample of 
Magellanic Cloud O stars using CMFGEN.  
Crowther et al.\ (2002) studied four extreme O supergiants (luminosity class
``Iaf+") in the Clouds, and derived temperatures that are considerably lower
than those shown 
in Fig.~\ref{fig:results}a 
for our sample. Hillier et al.\ (2003) analyzed the SMC
stars extreme supergiant Az~83 (O7 Iaf+) and Az 69 (OC7.5 III((f))).
The supergiant is also cooler than what we find, but this may be due to
the fact that the more extreme supergiants have lower effective temperatures.
The giant star agrees well with the Galactic giants.
Bouret et al.\ (2003) have examined the spectra of five
O dwarfs and one O-type giant in the NGC~346 cluster in the SMC. Here
the results are more mixed. The 
temperature of the star they classify as an O2~III is in accord with what
we find (Fig.~25b), as are some of the results for the dwarfs.  Others
of their dwarfs have effective temperatures lower than what Repolust
et al.\ (2004) found for Galactic stars of the same spectral type.
Are the differences due to the different models being
employed? We withhold any judgement until a similar sized sample of stars
have been analyzed with both codes.  

In the second paper in this series we will complete our Magellanic Cloud
sample.  At that time, we will also examine the effect that the new
data have on the wind-momentum luminosity relation, and give a comparison
between the masses derived from the present atmosphere fits and those of
stellar evolutionary models.

\acknowledgments

This paper draws heavily upon the {\it HST} archive, and it 
thus seems appropriate acknowledge the ease and 
convenience of a valuable resource 
often taken for granted in our community.  We are also very grateful to 
our program coordinator Beth Perriello, without whose efforts
our new {\it HST} data could not have been obtained.
Support for programs GO-6416, GO-7739, GO-8633, and GO-9412 was provided by
NASA through grants from the Space Telescope Science Institute, 
which is operated by the Association of Universities for Research in 
Astronomy, Inc., under NASA contract NAS 5-26555.
Useful comments on an earlier draft of this paper were kindly given by
Peter Conti, Artemio Herrero, Deidre Hunter, Nancy Morrison,
Daniel Schaerer, and Bill Vacca.
Nolan Walborn waived anonymity as the referee and made a large number of 
suggestions, for which we are grateful.
We also acknowledge the excellent support received during our observing
run at Cerro Tololo in January 1999, and in particular the efforts of
our telescope operator Patricio Ugarte.

\clearpage
\begin{deluxetable}{l l l l c c c c c l}
\pagestyle{empty}
\rotate
\tabletypesize{\scriptsize}
\tablewidth{0pc}
\tablenum{1}
\tablecolumns{10}
\tablecaption{\label{tab:stars}Program Stars\tablenotemark{a}}
\tablehead{
\colhead{Name\tablenotemark{b}}
&\colhead{Cat ID\tablenotemark{c}}
&\colhead{$\alpha_{2000}$}
&\colhead{$\delta_{2000}$}
&\colhead{$V$}
&\colhead{$B-V$}
&\colhead{$U-B$}
&\colhead{$E(B-V)$\tablenotemark{d}}
&\colhead{$M_V$\tablenotemark{e}}
&\colhead{Spectral Type\tablenotemark{f}}
}
\startdata 
AV 14                                 &SMC-007187&00 46 32.57&  -73 06 05.4&13.55&-0.17&-1.00&0.15&-5.8&O5~V    \\
AV 26                                 &SMC-009337&00 47 49.99&  -73 08 20.7&12.46&-0.17&-1.00&0.16&-7.0&O6~I(f)\\
AV 75                                 &SMC-016828&00 50 32.39&  -72 52 36.1&12.70&-0.15&-1.00&0.20&-6.9&O5.5 I(f+)\\
AV 207                                &SMC-043724&00 58 33.20&  -71 55 46.8&14.25&-0.20&-1.06&0.11&-5.0&O7.5 V((f))\\
AV 296                                &SMC-052948&01 02 08.57&  -72 13 19.9&14.26&-0.19&-1.02&0.13&-5.1&O7.5 V((f))\\
AV 372                                &SMC-055537&01 03 10.49&  -72 02 14.2&13.03&-0.16&-1.04&0.15&-6.8&O9.5~I\\
AV 377                                &SMC-061105&01 05 07.42&  -72 48 18.1&14.45&-0.20&-1.06&0.14&-4.9&O5~V((f))\\
AV 378                                &SMC-061202&01 05 09.39&  -72 05 34.7&13.77&-0.20&-1.02&0.12&-5.5&O9.5~III\\
AV 396                                &SMC-063226&01 06 04.25&  -72 13 34.2&14.10&-0.19&-1.00&0.12&-5.2&B0~III\\
AV 451                                &SMC-071002&01 10 26.06&  -72 23 28.9&13.97&-0.22&-1.04&0.07&-5.2&O9.5~III\\
AV 469                                &SMC-073337&01 12 28.97&  -72 29 29.2&13.12&-0.16&-1.05&0.13&-6.2&O8.5~I(f)\\
LH64-16=LH64:W16-8                         &LMC-142269&05 28 46.97&  -68 47 47.9&13.61&-0.22&-1.11&0.11&-5.2&ON2~III(f*)\tablenotemark{g}\\
LH81:W28-5\tablenotemark{h}       &  \nodata &05 34 28.47&  -69 43 56.6&13.92&-0.18&-1.10&0.15&-5.0&O4~V((f+))\\
LH101:W3-24=ST5-27\tablenotemark{i}& \nodata &05 39 14.10&  -69 30 03.8&14.58&-0.10&-1.00&0.25&-4.7&O3~V((f))\\
R136-020=R136a5\tablenotemark{j}      &  \nodata &05 38 42.5208&-69 06 03.112 & 13.93&\nodata&\nodata&0.42 & -6.0 & O2~If*\\
R136-024=R136a7\tablenotemark{j}      &  \nodata &05 38 42.4992&-69 06 02.961 & 14.06&\nodata&\nodata&0.41 & -5.8 & O3~III(f*)\\
R136-036=R136a-608\tablenotemark{j}   &  \nodata &05 38 42.7584&-69 06 03.214 & 14.49&\nodata&\nodata&0.46 & -5.5 & O2~If*\tablenotemark{k}\\
R136-040=R136a-535\tablenotemark{j}   &  \nodata &05 38 42.5555&-69 06 01.587 & 14.60&\nodata&\nodata&0.38 & -5,2 & O3~V\\
R136-047=R136a-602\tablenotemark{j}   &  \nodata &05 38 42.7331&-69 06 03.658 & 14.68&\nodata&\nodata&0.46 & -5.3 & O2~III(f*)\\
R136-055=R136a-551\tablenotemark{j}   &  \nodata &05 38 42.5984&-69 06 04.977 & 14.86&\nodata&\nodata&0.40 & -5.0 & O3~V\\
\enddata
\tablenotetext{a}{Coordinates and photometry 
are from Massey 2002, unless otherwise noted.}
\tablenotetext{b}{Identifications are as follows: 
``AV" from Azzopardi \& Vigneau 1982;
``LH" from Lucke 1972, 
``W" from Westerlund 1961;
``ST" from Testor \& Niemela 1998;
``R136-NNN" from Hunter et al.\ 1997 and Massey \& Hunter 1998;
``R136aN" from Weigelt \& Baier 1985,
``R136a-NNN" from Malumuth \& Heap 1994.}
\tablenotetext{c}{Designations from the catalog of Massey 2002.}
\tablenotetext{d}{From averaging the color excesses in 
$B-V$ and $U-B$ based upon the spectral type.  See Massey 1998b.}
\tablenotetext{e}{Computed using $A_V=3.1\times E(B-V)$, 
with assumed distance moduli for the SMC and LMC of 18.9 and 18.5,
respectively (Westerlund 1997, van den Bergh 2000).  The $M_V$ values
for the R136 stars were taken from Massey \& Hunter 1998.}
\tablenotetext{f}{New to this paper.}
\tablenotetext{g}{Classified as O2~III(f*) by Walborn et al.\ 2002, and
reclassifed as ON2~III(f*) by Walborn et al.\ 2004.}
\tablenotetext{h}{Coordinates and photometry from Massey, Waterhouse, \& 
DeGioia-Eastwood 2000.}
\tablenotetext{i}{Coordinates and photometry from Testor \& Niemela 1998.}
\tablenotetext{j}{Coordinates and photometry from Hunter et al.~1997
and Massey \& Hunter 1998. Although $UBV$ photometry has been
published by Malumuth \& Heap 1994, this was based upon pre-COSTAR
imaging; comparison with $V$ from the post-COSTAR data of Hunter et
al.~ 1997 shows poor agreement.}
\tablenotetext{k}{Classified as O2-O3~If* by Walborn et al.\ 2002, who refer to the star as ``MH 36"}
\end{deluxetable}

\clearpage
\begin{deluxetable}{l l l l c c c c c r}
\setlength{\tabcolsep}{0.02in}
\pagestyle{empty}
\tabletypesize{\scriptsize}
\tablewidth{0pc}
\tablenum{2}
\tablecolumns{10}
\tablecaption{\label{tab:journal}Data Used In This Study}
\tablehead{
\colhead{Star}
&\colhead{Telescope}
&\colhead{Instrument}
&\colhead{Observer}
&\colhead{Date}
&\colhead{Aperture}
&\colhead{Grating}
&\colhead{Wavelength}
&\colhead{Resolution}
&\colhead{Exp.~time} \\
\multicolumn{4}{c}{}
&\colhead{year-month-day}
&\colhead{(arcsec x arcsec)}
&\colhead{}
&\colhead{(\AA)}
&\colhead{(\AA)}
&\colhead{(sec)}
}
\startdata
\cutinhead{UV}
AV 14       &HST/5444&FOS &Robert&1994-09-24&3.7x3.7&G130H&1140-1600&1.0&480\\
AV 26       &HST/5444&FOS &Robert&1995-03-11&3.7x3.7&G130H&1140-1600&1.0&480\\
AV 26       &HST/9012&STIS&Massey&2002-07-31&0.2x0.2&G140L&1150-1740&0.9&445\\
AV 75       &HST/5444&FOS &Robert&1995-01-18&3.7x3.7&G130H&1140-1600&1.0&480\\
AV 207      &HST/5444&FOS &Robert&1994-10-21&3.7x3.7&G130H&1140-1600&1.0&480\\
AV 296      &HST/8633&STIS&Massey&2000-09-02&0.2x0.2&G140L&1150-1740&0.9&210\\
AV 372      &HST/5444&FOS &Robert&1994-11-16&3.7x3.7&G130H&1140-1600&1.0&480\\
AV 377      &HST/5444&FOS &Robert&1995-11-02&3.7x3.7&G130H&1140-1600&1.0&480\\
AV 378      &HST/5444&FOS &Robert&1995-03-08&3.7x3.7&G130H&1140-1600&1.0&480\\
AV 396      &HST/5444&FOS &Robert&1995-01-29&3.7x3.7&G130H&1140-1600&1.0&480\\
AV 451      &HST/5444&FOS &Robert&1995-03-11&3.7x3.7&G130H&1140-1600&1.0&480\\
AV 469      &HST/5444&FOS &Robert&1995-03-09&3.7x3.7&G130H&1140-1600&1.0&480\\
LH64-16     &HST/8633&STIS&Massey&2001-01-01&0.2x0.2&G140L&1150-1740&0.9&150\\
LH81:W28-5  &HST/8633&STIS&Massey&2000-11-13&0.2x0.2&G140L&1150-1740&0.9&120\\
LH101:W3-24 &HST/8633&STIS&Massey&2001-01-31&0.2x0.2&G140L&1150-1740&0.9&360\\
R136-020    &HST/5297&GHRS&Heap  &1994-04-03&0.2x0.2&G140L&1160-1450&0.6&1792\\
R136-020    &HST/5297&GHRS&Heap  &1994-04-02&0.2x0.2&G140L&1460-1750&0.6&7168\\
R136-024    &HST/6018&GHRS&Heap  &1996-02-03&0.2x0.2&G140L&1160-1450&0.6&1795\\
R136-024    &HST/6018&GHRS&Heap  &1996-02-03&0.2x0.2&G140L&1460-1750&0.6&3590\\
R136-036    &HST/6018&GHRS&Heap  &1996-02-02&0.2x0.2&G140L&1160-1450&0.6&2992 \\
R136-036    &HST/6018&GHRS&Heap  &1996-02-02&0.2x0.2&G140L&1460-1750&0.6&5984\\
R136-040    &HST/6018&GHRS&Heap  &1996-02-03&0.2x0.2&G140L&1160-1450&0.6&3617 \\
R136-040    &HST/6018&GHRS&Heap  &1996-02-03&0.2x0.2&G140L&1460-1750&0.6&7181\\
R136-047    &HST/6018&GHRS&Heap  &1996-02-01&0.2x0.2&G140L&1160-1450&0.6&4189\\
R136-047    &HST/6018&GHRS&Heap  &1996-02-01&0.2x0.2&G140L&1460-1750&0.6&7779\\
R136-055    &HST/6018&GHRS&Heap  &1996-02-02&0.2x0.2&G140L&1160-1450&0.6&9792\\
R136-055    &HST/6018&GHRS&Heap  &1996-02-02&0.2x0.2&G140L&1460-1750&0.6&4189\\
\cutinhead{H$\alpha$}
AV14       &CTIO 4-m& RC  &Massey&1999-01-06&1.3x330&KPGLD&5400-7800&2.8&600\\
AV26       &CTIO 4-m& RC  &Massey&1999-01-06&1.3x330&KPGLD&5400-7800&2.8&300\\
AV26       &HST/9412&STIS &Massey&2002-07-31&0.2x52 &G750M&6300-6850&0.8&360\\
AV75       &CTIO 4-m& RC  &Massey&1999-01-06&1.3x330&KPGLD&5400-7800&2.8&300\\
AV207      &CTIO 4-m& RC  &Massey&1999-01-06&1.3x330&KPGLD&5400-7800&2.8&376\\
AV296      &CTIO 4-m& RC  &Massey&1999-01-06&1.3x330&KPGLD&5400-7800&2.8&900\\
AV372      &CTIO 4-m& RC  &Massey&1999-01-06&1.3x330&KPGLD&5400-7800&2.8&300\\
AV377      &CTIO 4-m& RC  &Massey&1999-01-06&1.3x330&KPGLD&5400-7800&2.8&1200\\
AV469      &CTIO 4-m& RC  &Massey&1999-01-06&1.3x330&KPGLD&5400-7800&2.8&300\\
LH64-16    &CTIO 4-m& RC  &Massey&1999-01-06&1.3x330&KPGLD&5400-7800&2.8&600\\
LH81:W28-5 &CTIO 4-m& RC  &Massey&1999-01-06&1.3x330&KPGLD&5400-7800&2.8&600\\
LH101:W3-24&HST/9412&STIS &Massey&2002-05-30&0.2x52 &G750M&6300-6850&0.8&1800\\
R136-020   &HST/7739&STIS &Massey&1998-02-04&0.2x52 &G750M&6300-6850&0.8&280\\
R136-024   &HST/7739&STIS &Massey&1998-02-03&0.2x52 &G750M&6300-6850&0.8&370\\
R136-036   &HST/7739&STIS &Massey&1998-02-04&0.2x52 &G750M&6300-6850&0.8&490\\
R136-040   &HST/7739&STIS &Massey&1998-02-04&0.2x52 &G750M&6300-6850&0.8&280\\
R136-047   &HST/7739&STIS &Massey&1998-02-05&0.2x52 &G750M&6300-6850&0.8&660\\
R136-055   &HST/7739&STIS &Massey&1998-02-03&0.2x52 &G750M&6300-6850&0.8&775\\
\cutinhead{Blue}
AV14       &CTIO 4-m& RC  &Massey&1999-01-03&1.3x330&KPGLD&3750-4900&1.4&900\\
AV26       &CTIO 4-m& RC  &Massey&1999-01-04&1.3x330&KPGLD&3750-4900&1.4&600\\
AV75       &CTIO 4-m& RC  &Massey&1999-01-05&1.3x330&KPGLD&3750-4900&1.4&900\\
AV207      &CTIO 4-m& RC  &Massey&1999-01-04&1.3x330&KPGLD&3750-4900&1.4&600\\
AV296      &CTIO 4-m& RC  &Massey&1999-01-05&1.3x330&KPGLD&3750-4900&1.4&900\\
AV372      &CTIO 4-m& RC  &Massey&1999-01-05&1.3x330&KPGLD&3750-4900&1.4&300\\
AV377      &CTIO 4-m& RC  &Massey&1999-01-04&1.3x330&KPGLD&3750-4900&1.4&600\\
AV378      &CTIO 4-m& RC  &Massey&1999-01-04&1.3x330&KPGLD&3750-4900&1.4&747\\
AV396      &CTIO 4-m& RC  &Massey&1999-01-05&1.3x330&KPGLD&3750-4900&1.4&600\\
AV451      &CTIO 4-m& RC  &Massey&1999-01-04&1.3x330&KPGLD&3750-4900&1.4&600\\
AV469      &CTIO 4-m& RC  &Massey&1999-01-05&1.3x330&KPGLD&3750-4900&1.4&600\\
LH64-16    &CTIO 4-m& RC  &Massey&1999-01-04&1.3x330&KPGLD&3750-4900&1.4&1200\\
LH81:W28-5 &CTIO 4-m& RC  &Massey&1999-01-07&1.3x330&KPGLD&3750-4900&1.4&300\\
LH101:W3-24&CTIO 4-m& RC  &Massey&1999-01-07&1.3x330&KPGLD&3750-4900&1.4&542\\
R136-020   &HST/7739&STIS &Massey&1998-02-04&0.2x52& G430M&4310-4590&0.4&650\\
R136-020   &HST/6417&FOS  &Massey&1996-11-12&0.26(circ)&G400&3250-4820&3.0&770\\
R136-024   &HST/7739&STIS &Massey&1998-02-03&0.2x52& G430M&4310-4590&0.4&825\\
R136-024   &HST/6417&FOS  &Massey&1996-11-12&0.26(circ)&G400&3250-4820&3.0&870\\
R136-036   &HST/7739&STIS &Massey&1998-02-04&0.2x52& G430M&4310-4590&0.4&1150\\
R136-036   &HST/6417&FOS  &Massey&1996-11-13&0.26(circ)&G400&3250-4820&3.0&994\\
R136-040   &HST/7739&STIS &Massey&1998-02-04&0.2x52& G430M&4310-4590&0.4&1330\\
R136-040   &HST/6417&FOS  &Massey&1997-01-01&0.26(circ)&G400&3250-4820&3.0&1042\\
R136-047   &HST/7739&STIS &Massey&1998-02-05&0.2x52& G430M&4310-4590&0.4&1550\\
R136-047   &HST/6417&FOS  &Massey&1996-11-14&0.26(circ)&G400&3250-4820&3.0&1118\\
R136-055   &HST/7739&STIS &Massey&1998-02-03&0.2x52& G430M&4310-4590&0.4&1750\\
R136-055   &HST/6417&FOS  &Massey&1996-11-14&0.26(circ)&G400&3250-4820&3.0&1180\\
\cutinhead{Near-UV}
AV26       &HST/9412&STIS &Massey&2002-07-31&0.2x52& G430M&3020-3300&0.4&250\\
LH101:W3-24&HST/9412&STIS &Massey&2002-05-30&0.2x52& G430M&3020-3300&0.4&200\\
\enddata
\end{deluxetable}

\clearpage
\begin{deluxetable}{l c l}
\pagestyle{empty}
\tabletypesize{\footnotesize}
\tablewidth{0pc}
\tablenum{3}
\tablecolumns{3}
\tablecaption{\label{tab:termvels}Terminal Velocities in km s$^{-1}$}
\tablehead{
\colhead{Star}
&\colhead{$v_\infty$}
&\colhead{Comments\tablenotemark{a}}
}
\startdata
AV 14       & 2000: & Weak emission \\
AV 26       & 2150  & Second measurement: 2150 \\
AV 75       & 2100 & \\
AV 207      & \nodata & No measurable wind\\
AV 296      & 2000  & \\
AV 372      & 2000  &   C IV and Si IV\\
AV 377      & 2350:: & From N~V alone\\
AV 378      & \nodata & No measurable wind\\
AV 396      & \nodata & No measurable wind\\
AV 451      & \nodata & No measurable wind\\
AV 469      & 2000    & C IV and Si IV\\
LH64-16     & 3250    & \\
LH81:W28-5  & 2700    & \\
LH101:W3-24 & 2400    & \\
R136-020    & 3400 & dHH: 3000 \\
R136-024    & 3100 & P\&C: $_{\rm edge}=$3135, dHH: 2900 \\
R136-036    & 3700 & P\&C: 3640; dHH: 3750 \\
R136-040    & 3400 & P\&C: 3000; dHH: 3400 \\
R136-047    & 3500 & P\&C: 3305; dHH: 3625 \\
R136-055    & 3250 & P\&C: 2955; dHH: 3150 \\
\enddata
\tablenotetext{a}
{P\&C=Prinja \& Crowther 1998; dHH=de Koter, Heap, \& Hubeny 1998}
\end{deluxetable}

\clearpage
\begin{deluxetable}{l l c c c c c c r c c c r l}
\pagestyle{empty}
\rotate
\tabletypesize{\scriptsize}
\tablewidth{0pc}
\tablenum{4}
\tablecolumns{14}
\tablecaption{\label{tab:results}Results of Model Fits}
\tablehead{
\colhead{Name}
&\colhead{Spectral}
&\colhead{$T_{\rm eff}$}
&\colhead{$\log g$}
&\colhead{$R$}
&\colhead{$M_V$}
&\colhead{BC}
&\colhead{$M_{\rm bol}$}
&\colhead{Mass}
&\colhead{M-dot}
&\colhead{$\beta$}
&\colhead{$v_\infty$}
&\colhead{He/H}
&\colhead{Comments} \\
\colhead{}
&\colhead{Type}
&\colhead{($^\circ$K)}
&\colhead{[cgs]}
&\colhead{($R_\odot$)}
&\colhead{mags}
&\colhead{mags}
&\colhead{mags}
&\colhead{$M_\odot$}
&\colhead{($10^{-6} M_\odot$ yr$^{-1}$)}
&\colhead{}
&\colhead{(km s$^{-1}$)}
&\colhead{(by number)}
&\colhead{}

}
\startdata 
AV 14      &O5~V        & 44000 &    4.0&   14.2& -5.8 & -4.04 & -9.8  & 74& 0.1:&    0.8&   2000&    0.10& $\dot{M}<~0.3$   \\
AV 26      &O6~I(f)   & 38000 &    3.5&   27.5& -7.0 &  -3.63 & -10.6  & 87& 2.5&    0.8&   2150&    0.10&   \\
AV 75      &O5.5~I(f)  & 40000 &   3.6&   25.4& -6.9 &  -3.79 & -10.7  &94& 3.5&    0.8&   2100&    0.10&    \\
AV 207     &O7.5~V((f)) & 37000 &  3.7&   11.0& -5.0 &  -3.53 & -8.6   & 22& 0.1:&    0.8&   2000&    0.10& $\dot{M}<0.3$    \\
AV 296     &O7.5~V((f)) & 35000 &  3.5&   11.9& -5.1 &  -3.36 & -8.5   & 16& 0.5&    0.8&   2000&   0.10 &Poor fit. Binary?   \\
AV 372     &See Text    &\nodata&\nodata& \nodata& -6.8 &\nodata&\nodata& \nodata &\nodata& \nodata & 2000 & \nodata & Binary \\
AV 377     &O5~V((f))   & 45500 &    4.0&    9.1& -4.9 & -4.14& -9.1   & 30&    0.1:&    0.8&   2350&    0.35& $\dot{M}<0.3;$ Strong NIII\\
AV 378     &O9.5~III    & 31500 &  3.25 &   15.4& -5.5 & -3.06& -8.6   & 15&   (0.1)   & (0.8) & (2000) & 0.15& $\dot{M}$ not fit \\
AV 396     &B0~III    & 30000 &  3.5  &   14.1& -5.2 & -2.96& -8.2  & 23&    (0.1)  &  (0.8) & (2000) &  0.15& $\dot{M}$ not fit\\
AV 451     &See Text    &\nodata&\nodata& \nodata& -5.2 &\nodata&\nodata& \nodata& \nodata& \nodata & \nodata & \nodata & Binary \\
AV 469     &O8.5~I(f)   & 32000 &    3.1&   21.2& -6.2 & -3.12 & -9.3 & 21&    1.8&    0.8&   2000&    0.20&  \\
LH64-16    &ON2 III(f*)  & 54500 &   3.9&    9.4& -5.2 & -4.67 & -9.9 & 26&   4.0&    0.8&   3250&    1.0&See text     \\
LH81:W28-5 &O4~V((f+))   & 46000 &    3.8&    9.6& -5.0 & -4.17 &-9.2 & 21&   1.2&    0.8&   2700&    0.20&     \\
LH101:W3-24&O3~V((f))   & 48000  &    4.0&    8.1& -4.7 & -4.29 &-9.0 & 24&   0.5&    0.8&   2400&    0.15&    \\
R136-020   &O2~If*& $>$42500 &    3.6&   $>$16.4& -6.0 & $<$-3.98&$<$-10.0&$>$39&   23.0&    0.8&   3400&    0.20&$T_{\rm eff}$ lower limit\\
R136-024   &O3~III(f*)  &\nodata&\nodata&\nodata&-5.8&\nodata&\nodata& \nodata&\nodata&\nodata&\nodata&\nodata&Binary\\
R136-036   &O2~If*      & $>$43000 &    3.7&   $>$12.8& -5.5 & $<$-4.00 & $<$-9.5 & $>$30&   14.0&    0.8&   3700&    0.10& $T_{\rm eff}$ lower limit      \\
R136-040   &O3~V        & $>$51000 &    3.8&   $>$10.3& -5.3 & $<$-4.48  & $<$-9.8 &$>$24&    2.0&    0.8&   3400&    0.20& $T_{\rm eff}$ lower limit       \\
R136-047   &O2~III(f*)  & $>$51000 &    3.9&   $>$10.4& -5.3 & $<$-4.49  & $<$-9.8 &$>$32&  6.0&    0.8&   3500&    0.10& $T_{\rm eff}$ lower limit\\
R136-055   &O3~V        & 47500 &    3.8&   9.4& -5.0 & -4.26 & -9.3 & 20&   0.9&    0.8&   3250&    0.10&        \\
\enddata
\end{deluxetable}

\clearpage
\begin{figure}
\caption{\label{fig:diff} The difference between the standard
CALSTIS pipeline reduction (upper) and the re-reduced spectrum
with IRAF (lower) is shown for one of our program stars,
LH101:W3-24.  The data consist of a ``CR-SPLIT" 300~sec exposure
(i.e., two 150~sec segments) and are archived by STScI as data set
``o6km14010).  The complete
H$\alpha$ data set for this star consists of three such exposures.  The
SNR in the CALSTIS reduction is 18 in a spike-free region; the
same region in the IRAF-reduced spectrum has a SNR of 22.  The
many spikes (due to cosmic rays and hot pixels) in the CALSTIS spectrum
further degrade the SNR.}
\end{figure}


\begin{figure}
\end{figure}

\begin{figure}
\end{figure}

\begin{figure}

\caption{\label{fig:heap} A comparison of our spectra (green) with the
FOS spectra (red) used by de Koter et al.~(1998) for the six R136 stars.
In all cases there is additional emission present in the FOS data,
due to nebular contamination. 
}
\end{figure}

\begin{figure}

\caption{\label{fig:AV26comp} A comparison of our HST/STIS spectrum
(green) with the
ground-based CTIO spectra (red) of AV~26.}
\end{figure}

\begin{figure}

\caption{\label{fig:UV} The two observations of the C~IV $\lambda 1550$
profile are shown by the solid (Cycle 12) and dashed (Cycle 9)
black lines.  Our fit is shown
by the blue curve.  The fact that the fit does not reproduce the
absorption features on the longwards part of the profile
and the emission peak is due to there being no correction made for photospheric
absorption. This does not affect the determination of the
terminal velocity.} 

\end{figure}

\begin{figure}

\caption{\label{fig:AV14} AV 14.  (a) A portion of the 
blue-optical spectrum of AV~14 is shown with the major lines identified.
(b) Selected spectral lines (black) are shown compared to the model fits (red).
The bar to the left of each line shows a change of 20\% intensity relative
to the continuum, and the top of the bar denotes the continuum level. A radial velocity of 160  km s$^{-1}$ and a rotational
broadening $v \sin i$ of 150 km s$^{-1}$ were used in making this comparison.
}
\end{figure}

\begin{figure}
\caption{\label{fig:AV26} AV26. The same as Fig.~\ref{fig:AV14}, except that
a radial velocity of 120   km s$^{-1}$ and a rotational
broadening $v \sin i$ of 150 km s$^{-1}$ were used.
}

\end{figure}

\begin{figure}
\caption{\label{fig:AV75} AV75. The same as Fig.~\ref{fig:AV14}, except that
a radial velocity of 200 km s$^{-1}$ and a rotational broadening of
$v \sin i$ of 120 km s$^{-1}$ have been adopted.
}

\end{figure}


\begin{figure}
\caption{\label{fig:AV207} AV207. The same as Fig.~\ref{fig:AV14}, except that
a radial velocity of 95 km s$^{-1}$ and a rotational broadening of 
$v \sin i$ of 120 km s$^{-1}$ have been adopted.
}
\end{figure}

\begin{figure}
\epsscale{0.6}
\caption{\label{fig:AV296} AV296. The same as Fig.~\ref{fig:AV14}, except that
a radial velocity of 250 km s$^{-1}$ and a rotational broadening of 
$v \sin i$ of 300 km s$^{-1}$ have been adopted.
}

\end{figure}

\begin{figure}
\caption{\label{fig:AV372} AV372. This star is likely a double-lined
spectroscopic binary, with two stars contributing to the
He~I lines, and one star dominating the He~II lines.}
\end{figure}

\begin{figure}
\caption{\label{fig:AV377} AV377. The same as Fig.~\ref{fig:AV14}, except that
a radial velocity of 180 km s$^{-1}$ and a rotational broadening of
$v \sin i$ of 120 km s$^{-1}$ have been adopted.
}
\end{figure}


\begin{figure}
\caption{\label{fig:AV378} AV378.  The same as Fig.~\ref{fig:AV14}, except
that a radial velocity of 190  s$^{-1}$ and a rotational broadening of
$v \sin i$ of 110 km s$^{-1}$ have been adopted.  Note that the scale of
He~I$\lambda 4471$ has been changed with respect to earlier figures.}
\end{figure}

\begin{figure}
\caption{\label{fig:AV396} AV396.  The same as Fig.~\ref{fig:AV14}, except
that a radial velocity of 170  s$^{-1}$ and a rotational broadening of
$v \sin i$ of 120 km s$^{-1}$ have been adopted.  Note that the scale of
He~I$\lambda 4471$ has been changed with respect to Fig.~\ref{fig:AV14}.}
\end{figure}

\begin{figure}
\caption{\label{fig:AV451} AV451.  Like AV372 (Fig.~\ref{fig:AV372}), 
this star is judged to be
an incipiently resolved double-lined binary, with two stars contributing
He~I, and one star dominating the He~II contribution.  The composite
spectral type would be O9.5~III.
}

\end{figure}

\begin{figure}
\caption{\label{fig:AV469} AV469.   The same as Fig.~\ref{fig:AV14}, except
that a radial velocity of 180  s$^{-1}$ and a rotational broadening of
$v \sin i$ of 110 km s$^{-1}$ have been adopted.  Note that the scale of
He~I$\lambda 4471$ has been changed with respect to Fig.~\ref{fig:AV14}.}

\end{figure}

\begin{figure}
\caption{\label{fig:LH6416} LH64-16.   The same as Fig.~\ref{fig:AV14}, except
that a radial velocity of 290  s$^{-1}$ and a rotational broadening of
$v \sin i$ of 120 km s$^{-1}$ have been adopted.  The scale of 
He~I $\lambda 4471$ has now been magnified relative to that of earlier
figures.}

\end{figure}

\clearpage

\begin{figure}
\caption{\label{fig:LH81w5} LH81:W28-5.   
The same as Fig.~\ref{fig:AV14}, except
that a radial velocity of 280  s$^{-1}$ and a rotational broadening of
$v \sin i$ of 120 km s$^{-1}$ have been adopted.}

\end{figure}

\begin{figure}
\caption{\label{fig:lh101w24} LH101:W3-24.   
The same as Fig.~\ref{fig:AV14}, except
that a radial velocity of 280  s$^{-1}$ and a rotational broadening of
$v \sin i$ of 120 km s$^{-1}$ have been adopted. Note that a cosmetic
defect on the redward side of the He~II $\lambda 4200$ line, and
the nebular emission at H$\beta$, have been suppressed in making this figure. 
The H$\alpha$ profile is free of nebular contamination as it was obtained
with {\it HST}'s superior spatial resolution.}

\end{figure}

\begin{figure}
\caption{\label{fig:r136020} R136-020.  
In (a) we identify the major lines in the spectrum of R136-020.
The data are from the FOS observations of Massey \& Hunter (1998), except
for the region 4310\AA\ to 4590\AA, where we have spliced in our higher
SNR STIS spectrum.  The relatively high interstellar extinction in 
the 30~Dor region results in the strong diffuse interstellar band (DIB)
at $\lambda 4430$. In (b) we show the match between the model spectra
(red) and the observed spectrum (black).  As in earlier figures, the
bars denote a change in intensity of 20\%, and the tops of the bars
denote the continuum level. The He~II $\lambda 4686$ profile
is from the FOS data; the rest are from STIS. A rotational velocity of 120 km s$^{-1}$ has been adopted; uncertainties in the zero-point of the STIS
wavelength scale prevent accurate radial velocity measurements.
}
\end{figure}

\clearpage

\begin{figure}
\caption{\label{fig:r136024} R136-024.
The spectrum of R136-024 is likely a composite of two O3~V stars.
}
\end{figure}

\begin{figure}
\caption{\label{fig:r136036} R136-036.  The same as Fig.~\ref{fig:r136020}. 
}
\end{figure}

\begin{figure}
\caption{\label{fig:r136040} R136-040.  The same as Fig.~\ref{fig:r136020}.
}
\end{figure}

\begin{figure}
\caption{\label{fig:r136047} R136-047.  The same as Fig.~\ref{fig:r136020}.
}
\end{figure}

\begin{figure}
\caption{\label{fig:r136055} R136-055.  The same as Fig.~\ref{fig:r136020}. 
}
\end{figure}

\begin{figure}
\caption{\label{fig:3200} He~I$\lambda 3187$ and He~II $\lambda 3203$.
The near-UV region of the spectrum (black) is shown for AV~26 and LH101:W3-24,
along with the model fits (red).
}
\end{figure}

\clearpage

\begin{figure}
\caption{\label{fig:results} Effective temperature scale as a function of
spectral subtype.  The scale on the x-axis corresponds to the spectral
subtype (2=O2, 5.5=O5.5, 10=B0, 11=B1) for the three luminosity classes.
The black filled circles are the results of model fits to Galactic stars, taken
from Repolust et al.\ (2004). The red (LMC)  and green (SMC) 
filled circles are from
the present study, while the red (LMC) and green (SMC) triangles
are the results of CMFGEN modeling from Crowther et al.\ (2002),
Hillier et al.\ (2003), and
Bouret et al.\ (2003).
The solid line corresponds to the effective
temperature scale of Vacca et al.\ (1996), and the dashed line corresponds
to the effective temperature scale of Conti (1988).
}
\end{figure}

\end{document}